\documentclass[%
 jor,
 amsmath,amssymb,
 reprint,%
showkeys
]{revtex4-2}
\usepackage{graphicx}
\usepackage[version=4]{mhchem}
\usepackage{dcolumn}
\usepackage{bm}
\usepackage{siunitx}
\usepackage[hidelinks]{hyperref}
\usepackage{color}

\usepackage{afterpage}
\usepackage{hyperref}
\hypersetup{colorlinks,allcolors=black}
\begin{document}


\title[back-side ablation]{3D Free-Form Optical Lens - Miniaturised Fibre Couplers for Astrophotonics}

\author{
Haoran Mu$^{1,*}$, Daniel Smith$^{1,*}$, Tomas Katkus$^1$, Nguyen Hoai An Le$^1$, Dominyka Stonyte$^2$, Darius Gailevi\v{c}ius$^2$, Dan Kapsaskis$^1$, Alexander Del Frate$^{1}$,  Talwinder Singh Bedi$^3$, Donatas Narbutis$^4$,  Vijayakumar Anand$^5$, Darija Astrauskyte$^6$, Lina Grineviciute$^6$, Soon Hock Ng$^1$, Karl Glazebrook$^7$, Jon Lawrence$^8$, Saulius Juodkazis$^{1,2,9}$
}
\affiliation{$^1$ Optical Sciences Centre, ARC Training Centre in Surface Engineering for Advanced Materials (SEAM), Swinburne University of Technology, Hawthorn, Victoria 3122, Australia}
\affiliation{$^2$ Laser Research Center, Physics Faculty, Vilnius University, Saul\.{e}tekio Ave. 10, 10223 Vilnius, Lithuania}
\affiliation{$^3$ Chandigarh College of Engineering, Chandigarh Group of Colleges, Jhanjeri, Mohali - 140307, Punjab, India }
\affiliation{$^4$ Institute of Theoretical Physics and Astronomy, Faculty of Physics, Vilnius University, Sauletekio 9, 10222, Vilnius, Lithuania}
\affiliation{$^5$ Institute of Physics, University of Tartu, Estonia}
\affiliation{$^6$ Center for Physical Sciences and Technology, Savanoriu ave. 231, LT-02300,Vilnius, Lithuania}
\affiliation{$^7$ 
Centre for Astrophysics \& Supercomputing (CAS), Swinburne University of Technology, Victoria 3122, Australia}
\affiliation{$^8$ Australian Astronomical Optics (AAO), Macquarie University, NSW 2109, Australia}
\affiliation{$^9$ WRH Program International Research Frontiers Initiative (IRFI) Tokyo Institute of Technology, Nagatsuta-cho, Midori-ku, Yokohama, Kanagawa 226-8503 Japan}
\homepage{*H.M. and D.S. contributed equally.}
%

\date{\today}

\begin{abstract}
In astronomy, multi-object spectrographs employ fibre positioning robots to couple the light from multiple astronomy sources (stars or galaxies) into multiple multi-mode fibres, which are distributed across the focal plane of the telescope. These fibres transport the celestial light to the entrance slit of a spectrograph (or bank of spectrographs) for analysis. For any multi-object system mm-scale opto-mechanical solutions are required to couple the telescope light efficiently into the fibre.
We demonstrate a unique micro($\mu$)-optics solution to replace current optical fibre couplers. Specifically, we target technology on board the Keck telescope's FOBOS - Fibre-Optic Broadband Optical Spectrograph - which operates at UV to IR spectral ranges. For spectrally broad UV-IR band operation we use glass and crystals: fused silica, crystalline quartz (transparency $0.16 - 2~\mu$m), sapphire Al$_{2}$O$_{3}$ ($0.2 - 5~\mu$m), CaF$_2$ ($0.2-7~\mu$m), and BaF$_2$ ($0.2-10~\mu$m). The miniaturised $\mu$-coupler is monolithic, with the entire light path contained within glass or crystal, seamlessly extending to the fibre entrance, which is laser-machined and precisely aligned with the optical axis. 
\protect
      
\protect

\end{abstract}

\maketitle
\tableofcontents
\section{\label{Intro}Introduction}


\begin{figure*}[tb]
    \centering\includegraphics[width=15cm]{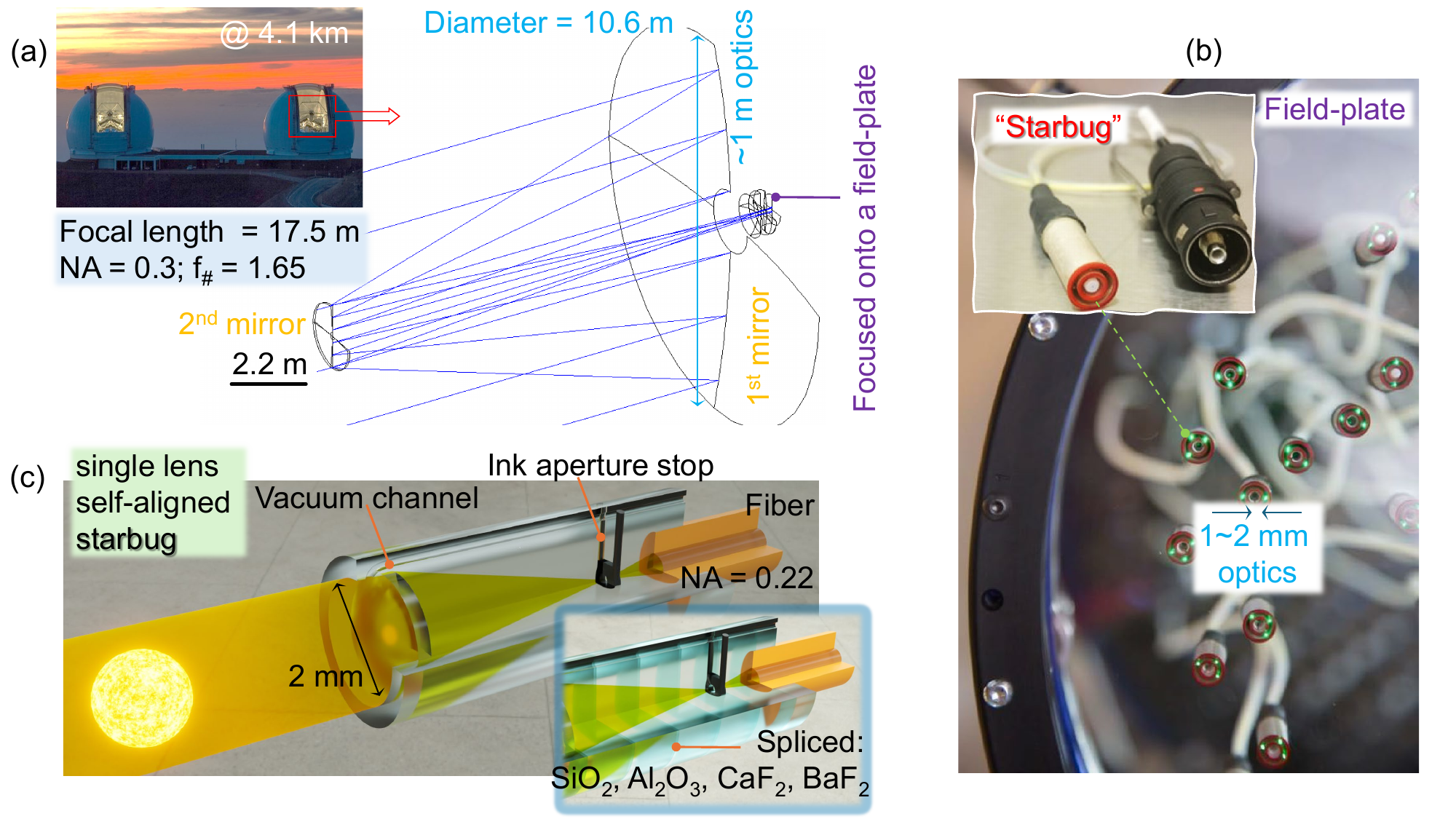} \caption{Micro-optics for astrophotonics. (a) Ray-tracing sketch (
    OSLO) for the Keck telescope with primary mirror of $D=10.6$~m and focal length $F=17.5$~m. Sky image is formed onto the telescope field-plane. Keck image: \url{https://www.keckobservatory.org/our-story/telescopes/}. 
 (b) Example of a fibre-coupled optical positioner with a piezo-actuator and pneumatic suction fixation to a field-plate at the telescope focal plane: Starbug. This system uses  a pair of free-space lenses aligned with the optical fibre for coupling. Image courtesy of~\cite{Lawrence_2016,foboS}. (c) Concept of a miniaturised free-form, single lens self-aligned optical fibre coupler: $\mu$-coupler. It could have integrated etched channels for pneumatic fixation, optical aperture stops. Single front surface delivers light collection matching the numerical aperture of the fibre $NA = 0.22$. Splicing of different materials for UV-VIS-IR functionality is amenable with fs-laser or \ce{CO2} laser micro-welding/joining.
    \label{f-keck}
    }
\end{figure*}

The development of new micro($\mu$)-optics-based photonic technologies has a proven record of success from lasers to fibre optics and telecommunications, especially in the 21$^{\textrm{st}}$ century where light 
drives industrial innovation. 
Spectral ranges from radio waves at centimetre scales to angstrom-level wavelengths (spanning approximately 10$^{9}$ orders of magnitude) are being increasingly explored and utilised, with particular emphasis on the visible and UV-IR regions.
Light is a unique source of information available to astronomers to probe the most distant objects and vast spaces. With larger developments of space technologies and their affordability, requirements for larger data transfer rates (bandwidth) evolve via the exploration of shorter wavelengths for telecommunications, considering even direct Earth-orbit links by optical channels. A miniaturised 0.5-mm-diameter high-power - continuous wave (cw) Photonic crystal surface-emitting lasers (PCSELs) at near-IR wavelengths have been developed delivering near-perfect Gaussian mode with quality factor $M^2 = 1$ at 1~W power for 10 Gbit/s high-speed transmission over 5~km long-distance without transmitter lens nor fibre amplifier~\cite{Morita}. The power scaling potentially up to kW-class of PCSELs via a surface area increase is feasible~\cite{Noda}. Miniaturisation and light weight are the same guiding principles for space exploration and satellite-based astronomy, which has a proven track record over the last 70 years in microelectronics, driven by exponential growth, the Moore's law~\cite{ASEC2021-11143}. 

For astronomical spectrographs, transmission across a wide UV-IR spectral range is important to capture a large amount of information about an object. The use of focusing $\mu$-optics made by direct 3D laser printing (additive or subtractive) are demanding due to the choice of materials, quality of surface definition, and control of its roughness. 
This is especially important in the case of coupling light into optical fibres (Fig.~\ref{f-keck}).
In astrophotonics, current fibre positioning devices use a fibre-coupled mm-scale opto-mechanical solution for coupling light from the telescope of interest efficiently into the optical fibres ~\cite{haynes2006wide,Bundy_2020,Saunders_2010,Staszak_2016,Kuehn_2014,Lawrence_2016,Goodwin_2015}. In this study, we aim to miniaturise the optics, based on a free-space assembly of two lenses that collect light into an optical fibre.

Femtosecond lasers enable flexible and almost thermal-damage-free subtractive ablation of solid materials~\cite{Joglekar2004}. However, debris accumulation and continuous surface shape changes during the ablation process can hinder laser focusing and prevent further ablation~\cite{23s2207968}. To address this, we employ a back-side ablation strategy, where the laser is focused through the substrate and focused onto the back-side surface to avoid scattering, back reflection as there is no $\pi$-phase shift due to light traversing from high-to-low index. All the debris which are in the laser beam are pushed along light propagation away from the surfaces by scattering and/or absorption (a transfer of linear momentum). Back-side ablation using liquid interface adds mode versatility to laser machining due to additional contribution of cavitation for material removal~\cite{23s2207968}. For fibre optical applications, fs-laser cut of optical fibres delivers superior end-flatness as compared with mechanical cleaving~\cite{21olt107111}. Also, optical edge or band-pass filters can be defined with high precision placement onto the fibre end using fs-fab~\cite{24n1345}. Uniquely for crystalline \ce{Al2O3}, fs-laser amorphised regions at the irradiation site can be removed by wet etching in KOH~\cite{08pssrrl275} or HF~\cite{06am1361} based aqueous solutions with very high contrast since crystalline phase has almost zero etch rate. The wet-etch assisted fs-laser fabrication at high concentrations of etchants and elevated temperatures $90^\circ-270^\circ$C approaching the boiling temperatures of the mixtures' constituents~\cite{etch} expands arsenal of approaches which can be used for fabrication of $\mu$-optical elements and fibre couplers.

Here we demonstrate different steps and methods for assembly of a fully functional micro-optical $\mu$-coupler to replace a pair of lenses in a current Starbug (Fig.~\ref{f-keck}b)). Starbugs are autonomous fibre positioning robots that were designed and tested on the TAIPAN instrument onboard the UK-Schmidt telescope in Siding Spring, NSW, Australia~\cite{Kuehn_2014,Staszak_2016,Bacigalupo_2018}, and are planned for use for Keck Observatory's planned FOBOS: the Fibre-Optic Broadband Optical Spectrograph~\cite{Lawrence_2016,foboS}. Such assemblies have a perfectly aligned fibre with the lens by design and are made from a single block (or spliced) from crystal or glass blocks into a single unit with a fibre port milled into it at the back-side. We use spectrally broad UV-IR band materials to enhance application range from astrophotonics to biomedical endoscopy using glass and crystals: fused silica, crystalline quartz, sapphire, CaF$_2$, and BaF$_2$. Design of the $\mu$-coupler is based on the '{e}tendue matching~\cite{etendue}, which is applied from the optical fibre (detection) side of an optical system, i.e., the numerical aperture (solid angle) and fibre diameter. Femtosecond (fs)-laser machining and etching, coating, annealing, joining assisted fabrication is explored. Practicality and simplicity of fabrication protocols was the primary goal aiming for prototyping $\mu$-couplers that could be used for Starbug fibre positioning robots for FOBOS, or other next generation fibre positioning systems in astronomy.     


\begin{figure}[tb!]
    \centering\includegraphics[width=7cm]{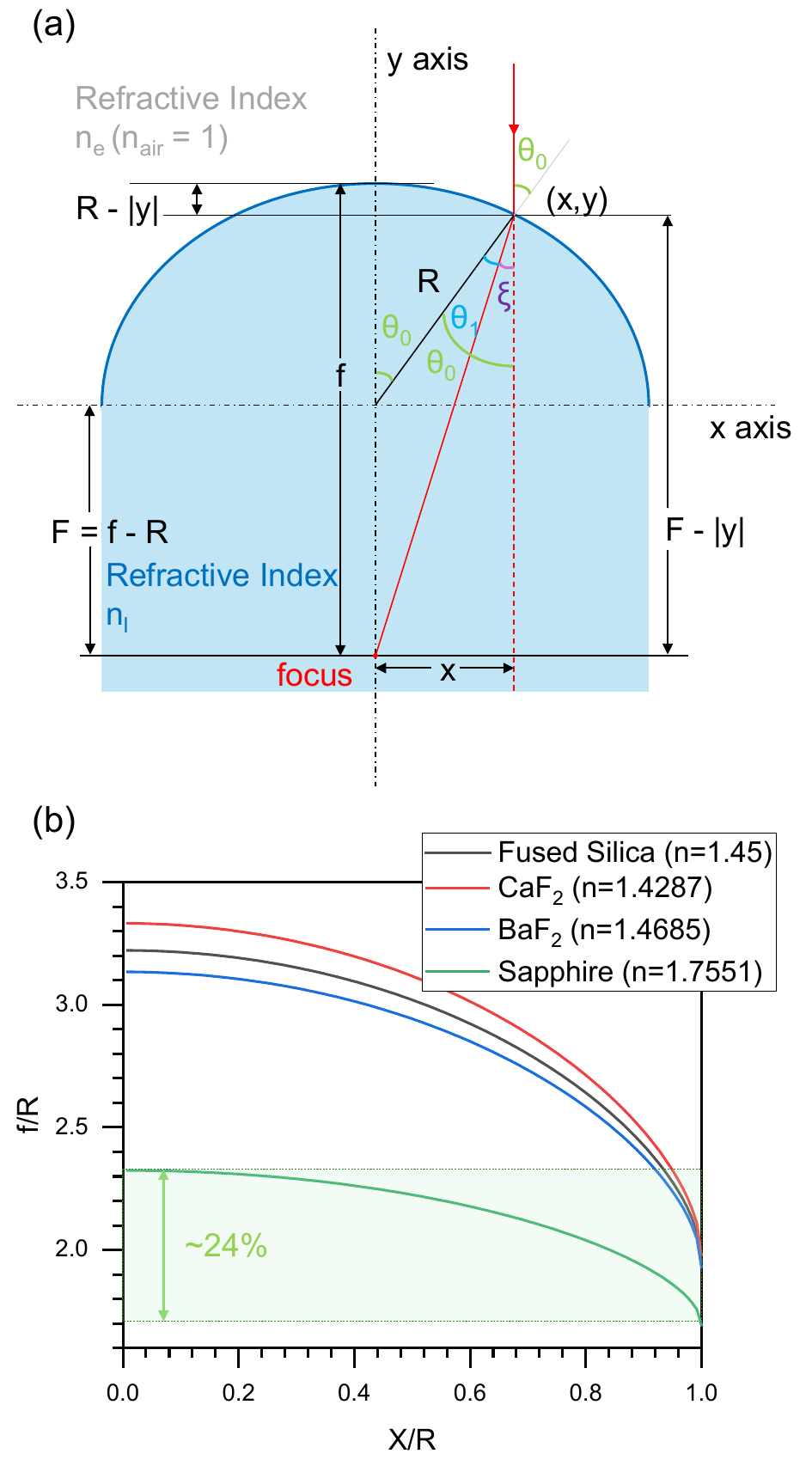}      \caption{\label{f-inside} (a) Analytical ray-optics model for a spherical surface focusing from medium $n_e$ (external) into lens material $n_l$; $n = n_l/n_e$. (b) Focal length from the convex side as a function of distance from centre normalised to the radius $R$: $f/R$ vs. $X/R$ for different materials $n$ (normalised index)~\cite{inside}.
}
\end{figure}


\begin{figure}[tb!]
    \centering\includegraphics[width=8.5cm]{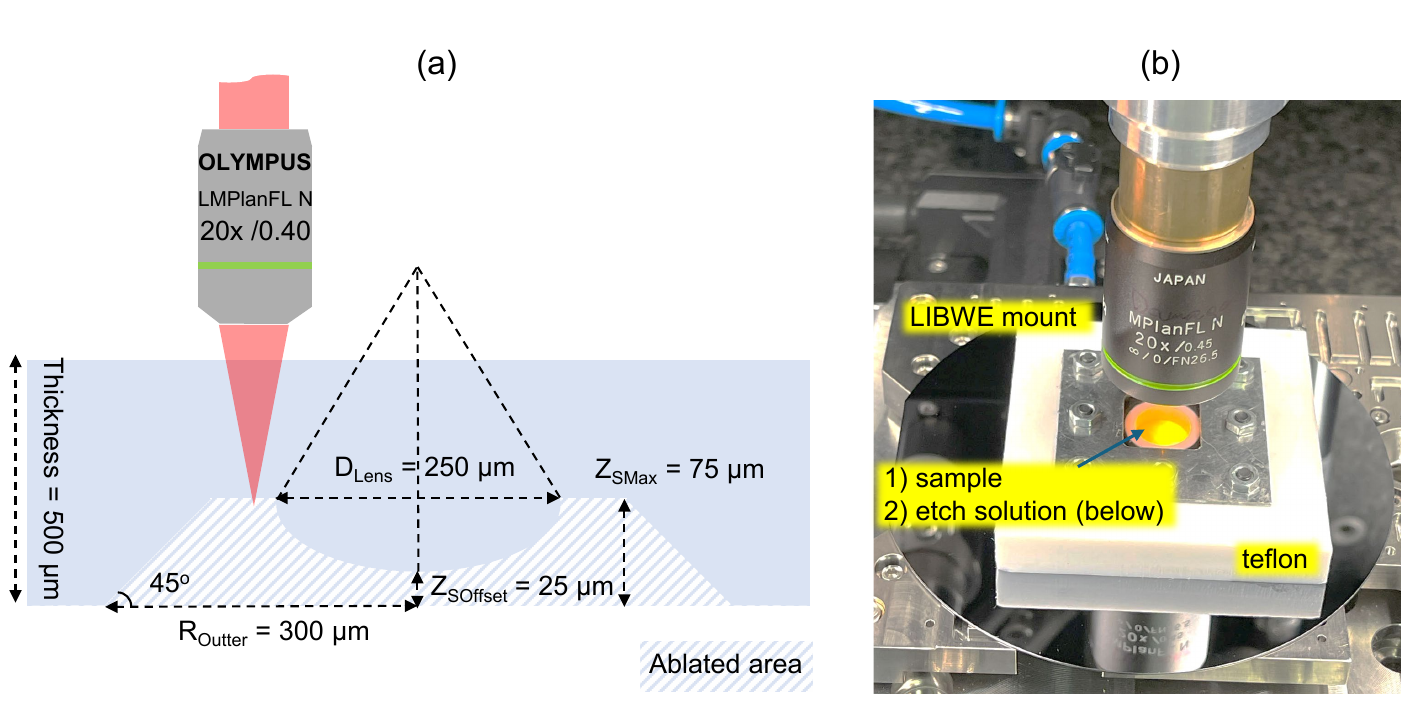}      \caption{\label{f-schematic} Back-side ablation in air (a) and liquid (b); LIBWE - light induced back-side wet etching. Samples of glasses and crystals were 0.5~mm. 
    A spherical lens was defined for $NA\approx 0.22$ focusing to match collection $NA$ of the optical fibre. Laser machining was with 1030~nm or 515~nm wavelength $\sim 200$~fs pulses focused with $NA\approx 0.4$ (if conditions differed it is explicitly stated). For LIBWE, carbon black (sumi) ink solutions in water and acetone was used. Geometry of a lens of 0.25~mm in diameter was used for process development for further up-scaling to 0.5~mm and final 2~mm diameters with the same laser exposure conditions. 
}
\end{figure}

\begin{figure}[tb!]
    \centering\includegraphics[width=7.5cm]{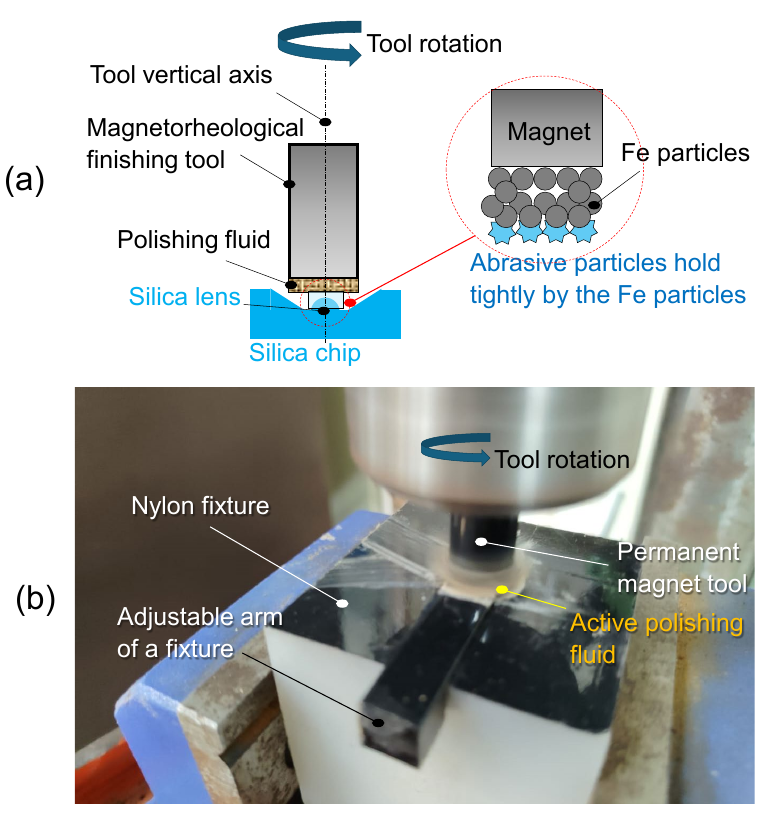}      \caption{\label{f-mrp}(a) Schematic representation of MRP - magnetorheological polishing. (b) Actual polishing of a silica $\mu$-lens with MRP tool. Polishing mixture: 20\%vol. Fe particles (grit 500), 10\%vol. Ce-oxide particles (grit 1000) and 70\%vol. deionised \ce{H2O}. 
}
\end{figure}

\section{Methods and materials}\label{Methods}

\noindent\textbf{Ray-optics design.} Figure~\ref{f-inside} shows geometry (a) and focusing performance (b) of a spherical surface when the focus is set inside the medium the lens is made of~\cite{inside}. This analytical model is based on simple expressions with focal length defined as  $f = \frac{x}{\tan\zeta}+\left( R-\sqrt{R^2 -x^2}\right)$ (Fig.~\ref{f-inside}(a)), where  $\sin\zeta = \frac{x}{R}\frac{1}{n}\left[\sqrt{n^2 - (x/R)^2} - \sqrt{1-(x/R)^2} \right]$, $\sin\theta_1 = (n_e/n_l)\sin\theta_0 = \frac{1}{n}\frac{x}{R}$; see geometrical notations in Fig.~\ref{f-inside}(a). The focal length defined by the lens maker's formula for two spherical surfaces with curvature radii $R_{1,2}$, respectively, and the lens thickness is $d$: $\frac{1}{f} = (n-1)\left[ \frac{1}{R_1} - \frac{1}{R_2} + \frac{(n-1)d}{nR_1R_2}\right]$; for our case with one spherical surface $R_2 = \infty$. It is insightful to present the normalised focal length (Fig.~\ref{f-inside}(b)) $\frac{f}{R} = \frac{x}{R}\frac{1}{\tan\zeta
} + 1 - \sqrt{1-(x/R)^2}$ as a function on position of the incoming ray $X/R$. It shows that a higher refractive index medium is beneficial to have a lesser axial extension of the focal region ($\sim 24\%$ marker in (b)) for the rays entering close to the aperture edge $x\approx \pm R$. Such spherical aberration can be eliminated by aspherical lens design, which was modelled for the $\mu$-coupler in this study. 

\noindent\textbf{UV-IR transparent crystals and glass.} Fused silica wafers (catalogue number 518, University Wafer Inc.), calcium fluoride substrates (\ce{CaF2} IR polished window 10$\times$0.5~mm, CAFP$10\times 0.5$) and barium fluoride substrates (\ce{BaF2} 10~mm $\times$ 0.5~mm polished window
BAFP10-0.5) were 0.5~mm-thick and were used for back-side laser machining. Two-side polished sapphire wafers (catalogue number 3490 University Wafer, Inc.) of 0.5~mm thickness and 76.2~mm in diameter were diced into smaller squares of 1~mm by 1~mm and were also used for back-side laser machining in air.
   
\noindent\textbf{Femtosecond laser machining.} Laser ablation with ultra-short pulses of 200~fs duration, 200.3~kHz repetition rate and the $1^{st}$ harmonic generation ($\lambda$ = 1030~nm) was used to 3D print the plano-convex front lens of the required f-number $f_\# = F/D = n/(2NA)\sim 3.85$ for the Keck telescope FOBOS spectrograph. Here $F = 5.5$~mm is the focal length inside material of refractive index $n\approx 1.4$ (for a monolithic lens-fibre coupler), $D = 2$~mm diameter of the front lens ($D =0.25,0.5$~mm were tested in this study for faster machining $\sim 15, 60$~min, respectively), $NA = n\sin\alpha$ is the equivalent numerical aperture matching $NA = 0.22$ of that of multi-mode optical fibre with core diameters $0.4~\mu$m or $0.2~\mu$m, which corresponds to the half-angle of light collection cone $\alpha\approx 10^\circ$. \\
The laser machining was carried out on the back-side (focusing through the sample onto the back-side) of 0.3-0.5~mm slabs in air. Single pulse fluence was typically up to ten times the ablation threshold in the range $15-25$~J/cm$^2$, average intensity (irradiance) $70-120$~TW/cm$^2$ at pulse density $\sim 750$~pulses/mm with pulse separation $\sim 1333.3$~nm (for in-line and between the lines in-plane and along the depth: slicing and hatching).
Beam was focused with an Olympus MPlanFL N 20$\times$, numerical aperture $NA = 0.45$ lens, focal spot diameter $d=1.22\lambda/NA = 2.8~\mu$m and depth-of-focus (DoF) or double the Rayleigh length $2z_R = 2\pi[d/2]^2/\lambda = 11.9~\mu$m, for $\lambda = 1030$~nm.

For ablation, the laser beam trajectory followed the lens profile and included a pedestal disk with an initial diameter of $2D$ (twice the lens diameter), which was reduced by 1.33~$\mu$m increments until it matched the lens diameter. For a lens with a diameter $D = 250~\mu$m and height $H = 50~\mu$m, the process was repeated 68 times with $800$~nm spacing steps along the depth. This also included a flat drilled depth $Z_{SOffset} = 25~\mu$m from the back-side surface to the top of the lens, adjustable based on application requirements. The stage movement speed was determined by a function involving the lens diameter and the maximum acceleration of the stage $v = \sqrt{D/2 \times A_{Max}}$, with a burst setting of 3, delivering three pulses simultaneously at each ablation site. This process sculptured an aspheric plano-convex lens profile and ablated an isosceles trapezoid shape with total depth $Z_{SMax} = 75~\mu$m and a $45^\circ$ slope angle for optimal mechanical support.

\noindent\textbf{Laser-Induced Back-side Wet Etching (LIBWE).} In our LIBWE realisation, carbon black (sumi) ink solutions in water and acetone were used. The sumi ink enhances laser energy absorption, increasing the temperature of fused silica to facilitate carbothermal reduction and prevent irregular crack propagation caused by the large thermal gradients between silica and acetone. The LIBWE chamber was made from Teflon using Computer Numerical Control (CNC) milling, chosen for its exceptional chemical resistance, including to solvents like acetone. An aluminium lid with eight screws was used to clamp the fused silica window to the Teflon chamber, sealing the liquid inside.

\noindent\textbf{Fs-laser welding/joining:} It was used to splice two silica slabs. The N-BK7 Plano-Convex lens (PART NO. LA1027) 
with  focal length = $35$~mm was used as practical tool for this task, where long focal region (DoF $\approx 200~\mu$m) is favourable. 
The laser beam, with a diameter of approximately $4$~mm, was focused by the spherical lens onto the interface of the pair, resulting in a focal spot diameter of about $11.45~\mu$m. 
Welding/joining was accomplished using the $200$~fs laser pulses with a repetition rate of $1$~MHz and a pulse density of $200$~pulses/$\mu$m, including a burst mode of 3 (three pulses delivered simultaneously). The two silica slabs were firmly compressed to eliminate air gaps and then taped onto a \ce{Si} wafer substrate, which was placed on the linear nanopositioning stages (ANT130XY, Aerotech) 
and secured using vacuum suction. To achieve strong welding/joining, the scanning speed was set to $0.5$~mm/s, the laser output power was $10$~W, delivering a single-pulse fluence of $9.7$~J/cm$^2$.

\noindent\textbf{Optical microscopy.} Structural characterisation and focal length measurements were conducted using an optical microscope (Nikon ECLIPSE LV100NPol). 
A 3D optical profiler (Bruker ContourGT InMotion) was employed to characterise the lens profile using a $NA = 0.45$ lens and to measure the average roughness $R_a$ of the ablated flat area with a $NA = 0.90$ lens. Scanning electron microscopy (SEM) was also used for structural characterisation, capturing a 45$^\circ$ image after sputtering a $20$~nm Cr coating. The Raith 150TWO electron beam writer in field-emission SEM mode was used for this analysis.

\noindent\textbf{Surface morphing with \ce{CO2} laser.} A \ce{CO2} laser (Epilog Helix 24) was employed to polish the fabricated fused silica lens. The process involved three key parameters: power, speed, and dots-per-inch (DPI). The maximum power is $75$~W, the maximum speed is $3$~mm/s, and DPI ranges from $75$ to $1200$. The system offers three job types: ``raster'' mode for engraving or marking, ``vector'' mode for cutting lines, and ``combined'' mode for simultaneous engraving and cutting. For polishing the fused silica lens, we used $20\%$ power $15$~W, $30\%$ speed $0.9$~mm/s, and the DPI setting of $1200$.

The diameter of \ce{CO2} cw-laser $\lambda = 10.6~\mu$m ($\Tilde{\nu} = 943$~cm$^{-1}$) at the focusing optics was $D = 3$~mm. The focal length was $F = 50$~mm, yielding an f-number $F_\# = F/D \approx 16.7$, which corresponds to a numerical aperture $NA = \frac{1}{2F_\#} \approx 0.03$. Consequently, the focal spot size on the sample, for a Gaussian beam with an ideal quality factor $M^2\equiv 1$, was determined to be  $2r = \frac{4\lambda}{\pi}\cdot\frac{F}{D} = 225~\mu$m. The axial extent of the focal region, known as the depth-of-focus DoF, was calculated to be
$DoF = \frac{8\lambda}{\pi}\cdot\left(\frac{F}{D}\right)^2 = 7528~\mu$m = 7.53~mm.

The energy deposition conditions are then characterised by the average intensity during laser writing, defined as $I = P/(\pi r^2)$~[W/cm$^2$]. For the scanning speed $v_{sc} = 0.9$~m/s the dwell time required to traverse the diameter of the focal spot was $t_{dw} = 2r/v_{sc} = 0.25$~ms.
The corresponding exposure dose $D_{ex}$ was obtained by multiplying $I$ and $t_{dw}$, resulting in $D_{ex} =  I\times t_{dw}$~[J/cm$^2$]; for $P = 15$~W, $I = 9.44$~kW/cm$^2$ the accumulation dose reached $D_{ex} = 0.59$~J/cm$^2$. In the case of a 2D raster scan and DPI value of 1200, there is a degree of overlap between neighbouring scanned lines making the overall 2D exposure dose of the pattern, $D_{2D} = \frac{DPI}{112}\cdot D_{ex} \approx 6.32$~[J/cm$^2$], where 112 represents the number of focal diameters per inch for a single line, and $D_{ex}$ corresponds to the dose for a single scan as defined previously.

\noindent\textbf{High Temperature Annealing (HTA).} HTA was employed using a Nabertherm Tube Furnace to smooth the surfaces of the fabricated optical elements by 
material reflow and sublimation~\cite{2011}. The annealing temperature was set at approximately 70-80\% of the material’s melting point to promote reflow while maintaining the shape of the lenses. The UV-IR transparent materials used, have high melting temperatures: CaF$_{2}$ at 1423$^{\circ}$C, quartz at 1670 - 1713$^{\circ}$C, and sapphire at 2050$^{\circ}$C. 

\noindent\textbf{Atomic Layer Deposition (ALD):} ALD~\cite{C2_S2_P1,C2_S2_P4,C2_S2_P6,C2_S2_P20} was used to deposit alumina Al$_{2}$O$_{3}$ films with thicknesses of 400~nm and 800~nm onto $\mu$-lenses 
fabricated on the c-plane of sapphire and underwent annealing for 2-4 hours at 1500$^{\circ}$C. The rationale for coating c-plane sapphire Al$_{2}$O$_{3}$ with alumina is that it ensures 
that the refractive index remains unchanged when light passes through the coated material.

In this research, depositions were carried out using a commercial ALD system Veeco Savannah S200. \ce{Al2O3} layers were deposited at 150$^\circ$C temperature, using trimethylaluminum (TMA) as a precursor and water as an oxidiser. Each ALD half-cycle consisted of 0.02~s TMA or \ce{H2O} exposure and 6~s \ce{N2} purge. The growth per cycle of aluminium oxide was 1.06\AA/c.

\noindent\textbf{Magnetorheological polishing (MRP):} The improvement in the surface quality of silica lens (0.25~mm) was done with the help of MRP technique\cite{MRP} (Fig.~\ref{f-mrp}). In MRP, a permanent magnet 
was fixed on the rotating chuck of a milling machine (model Mikrocut 2000S, Emtex). The dimensions of two permanent magnets (made of \ce{NdFe35}) used in this tool was 8~mm in diameter and 10~mm in height. 
The cube type nylon fixture was made 
by vertical milling center (VMC) (model CVM 1060, COSMOS). The silica chip was placed in the square fixture slot of $10\times 10$~mm$^2$ and it was further gripped with the help of fixture’s adjustable arm (Fig.~\ref{f-mrp}(b)). For experiment, a polishing fluid was 
mixed from different ingredients: 20~vol.\% electrolyte Fe powder having mesh size of 500 (minimum size 20~$\mu$m, maximum 40~$\mu$m), 10~vol.\% Ce-oxide abrasive powder having mesh size of 1000 (7 - 23~$\mu$m) and 70~vol.\% of de-ionised water as a base fluid. 
Under the magnetic environment, the fluid becomes active in nature, i.e. the Fe particles forms chain structures which become stuck to the magnet's surface, whereas abrasive particles because of their non-magnetic nature, were gripped by the Fe particles in the outward direction as schematically shown in Fig.~\ref{f-mrp}. This active fluid was used for performing a nano-scale finishing on the silica lens surface. 

The working gap between the magnet surface and silica lens was taken as 1~mm and thereafter a finishing was performed with tool rotational speed of 500~rpm. The surface roughness (initial and final finished) was measured with the help of 3D optical profilometry with the 115$\times$ $NA = 0.90$ objective lens. The initial $R_a$ value of silica lens was  $\sim 256$~nm. The MRP was performed on the $\mu$-lens surface for 180~minutes. After every 30~minutes of polishing, a fresh fluid was pasted on the magnet surface, so as to avoid the movement of the used abrasive particles inside the active polishing fluid. After overall finishing, the final $R_a$ value was found as $96.5\pm 6$~nm without a shape change (Fig.~\ref{f-magn}). 

\begin{figure*}[tb]
    \centering\includegraphics[width=16cm]{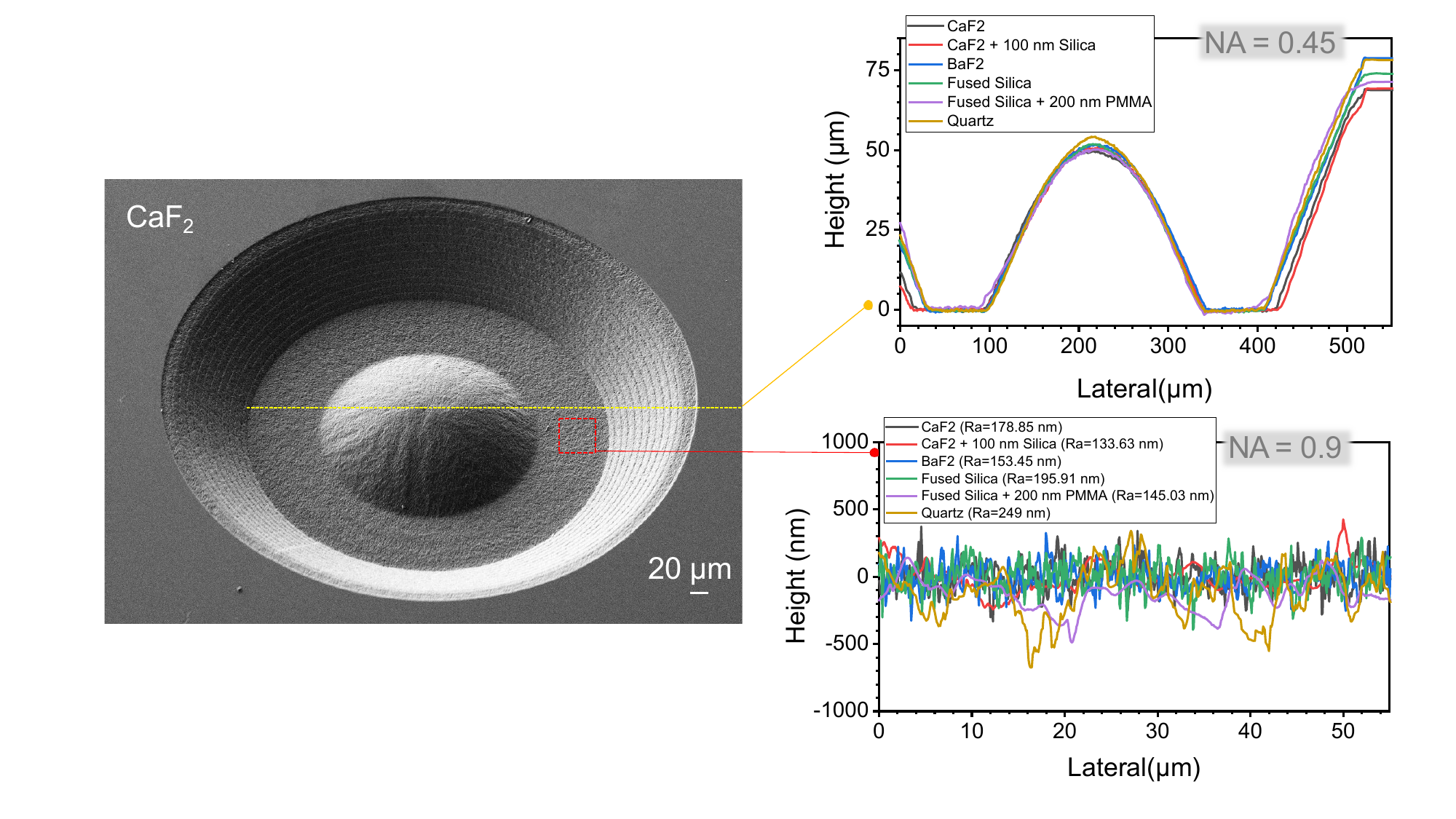}      \caption{\label{f-Ra} An average surface roughness of the back-side ablated spherical $\mu$-lenses made out of different materials: fused silica, quartz (001) plane, \ce{CaF2} and \ce{BaF2} with the same laser scanning protocol. Surface roughness was measured with different $NA$ objective lenses over the sphere-dome and on the flat region. After ablation e-beam evaporation of 100~nm silica and spin coating of 200~nm PMMA was used to reduce roughness $R_a$ (rms); thickness of spin coating corresponding to the plane coating. Width of the scan was one pixel and corresponded to 0.0890~$\mu$m ($115^\times$ 0.9$NA$ objective) and 0.9027~$\mu$m  $20^\times$ 0.45$NA$).  
}
\end{figure*}

\begin{figure}[tb]
    \centering\includegraphics[width=8.5cm]{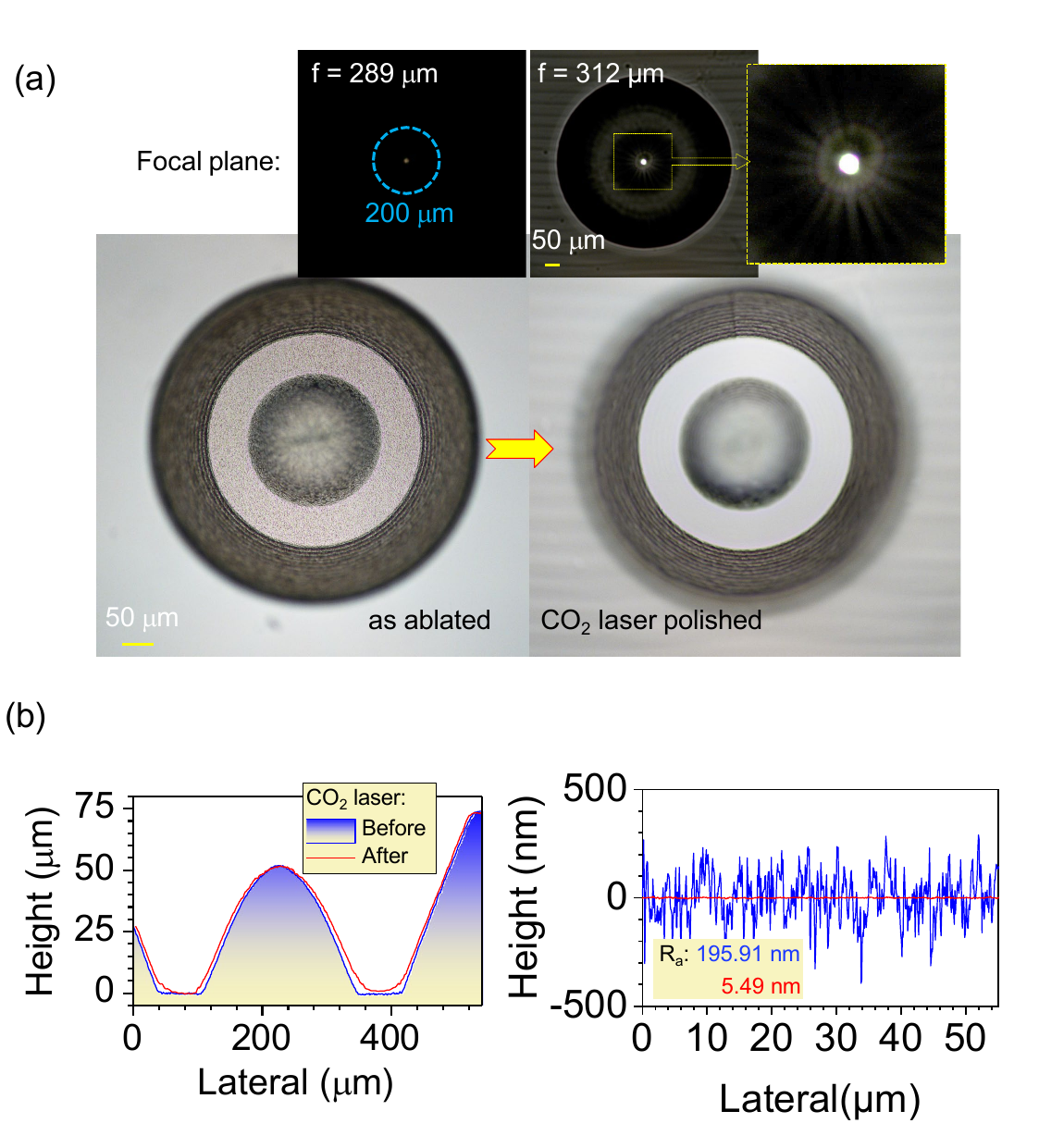}      \caption{\label{f-CO2} \ce{CO2} laser polishing of silica lens. (a) Optical image of the same lens before and after \ce{CO2} laser resurfacing. Smoothing condition: raster (horizontal) scan at 15~W, 0.9~m/s, and 1200 DPI. 
    Top-insets show images of the focal plane. The close-up view has a cross-section approximately of $200-\mu$m-diameter fibre core. (b) Cross sections after \ce{CO2} smoothing for the lens and flat region (as in Fig.~\ref{f-Ra}). Roughness (rms) $R_a\approx 5.5$~nm was achieved without detrimental shape morphing of the spherical micro-dome.  
}
\end{figure}

\section{Results} 

\subsection{Femtosecond laser 3D sculpturing}

In fs-laser ablation, the high 
intensity triggers nonlinear absorption processes, seeded by multiphoton absorption, avalanche ionisation, and partly by tunnelling ionisation at the tens-of-TW/cm$^2$, which lead to ionisation and free carrier plasma formation by direct molecular bond breaking~\cite{Joglekar2004}. Dielectric is transformed into metal-like medium where energy is deposited into shallow skin depth of tens-of-nm~\cite{DieMet}. This leads to energetic material removal with a minimal heat affected zone (HAZ). 
The rapid energy deposition 
limits heat diffusion and spread, significantly reducing thermal damage. This results in a highly precise and clean ablation process. 
However, not all material is vaporised or ejected from the surface with a plasma plume. A substantial amount of debris forms due to 
redeposition from plasma. This debris adheres to the machining area, disrupting the ablation conditions at the interface and causing scattering that hinders effective energy deposition and precise control over the ablation process. In $\mu$-optics fabrication, 
achieving a high optical surface finish is essential (surface roughness $<\lambda/15$), but debris accumulation and edge inhomogeneities present significant challenges to the process. To mitigate the impact of debris, a back-side ablation strategy was implemented. As shown in Fig.~\ref{f-schematic}, the laser beam passed through the transparent sample from the flat front side and was focused on the back-side. This approach allowed debris to be preferentially pushed along the propagation of laser pulse, markedly reducing  
debris formation. 

The lens profile and surface average roughness $R_a$ (rms) were measured using a 3D optical profiler (Bruker ContourGT InMotion) for $\mu$-lenses ablated on different materials with the same scanning pattern and optimised pulse energy aiming at small roughness. Figure~\ref{f-Ra} shows results of the lens profile  measured with a $NA = 0.45$ objective lens on different materials. The cross-sectional plot corresponds to the central cross section shown in the SEM image. The $R_a$ of the laser ablated flat surface region 
was measured using a higher resolution $NA = 0.90$ lens. The roughness of \ce{CaF2} was $\sim 178$~nm, 
\ce{BaF2} was $\sim 150$~nm, while that of silica glass $\sim 195$~nm or crystal $\sim 250$~nm as measured with the highest resolution on the flat region of ablated surface (lens profile has a signature of the scan geometry). For IR applications of \ce{CaF2} and \ce{BaF2}, such roughnesses correspond to the $\lambda/10 - \lambda/20$ and does not need further treatments even for the most demanding optical applications. For visible spectral range, roughness has to be significantly reduced to the acceptable $\lambda/10 \leq 40$~nm. By depositing 100~nm of silica by sputtering
, $R_a$ was reduced by comparable about on \ce{CaF2}. 
A spin coating of 200~nm of PMMA on silica had similar effect and reduced roughness of the ablated surface to 140~nm. These procedures are applicable and can be considered for applications, especially, when anti-reflection coatings are applied.       

\subsection{Laser-Induced Back-side Wet Etching (LIBWE)}

Figure~\ref{f-schematic} illustrates back-side ablation in two different environments: air (a) and liquid (b). The laser-induced back-side wet etching (LIBWE) was used with pure solvents (water and acetone) as well as water solution of black ink. 
In LIBWE, 
laser is focused onto the interface between the back-side surface of the transparent sample and the liquid. The deposited laser energy not only removes solid material but also results in the breakdown of the liquid into a high-temperature plasma, forming cavitation bubbles. These bubbles, which form at a lower breakdown threshold compared to solids, rapidly expand and contract, exerting fluctuating forces on the interface~\cite{23s2207968}. The energy deposited by the laser drives both the ablation and the cavitation processes. 
Cavitation induces the conversion of deposited laser energy into kinetic energy at the microscale, promoting the dynamic ejection of ablation products. This dynamic process continuously generates forces that push ablation debris away from the surface, allowing for clean and stable subtractive processing. As the ablation surface shape changes, the liquid refills the void created by previous laser pulses, ensuring continuous and effective material removal. Compared to back-side ablation in air, the roughness (rms) $R_a$ was reduced by $22.5\%$ when using LIBWE (Fig.~\ref{f-ink}). Further refinement of lens milling by LIBWE for a more efficient material removal rate as well as final surface roughness could be achieved via a choice of absorbing dyes and irradiation wavelength.    

\subsection{\texorpdfstring{\ce{CO2}}{TEXT} laser polishing}


To reduce the surface roughness, especially for UV applications, $R_a\approx\lambda/15\sim 20$~nm is required. For fs-laser ablated fused silica lens, we employed a \ce{CO2} laser (Epilog Helix 24) for surface polishing. Unlike crystalline materials, fused silica (glass) 
has a softening point at which it can be morphed. When the temperature exceeds this softening point, surface tension can reshape surface of silica by reflow~\cite{2011}. It can be effectively used to smooth the laser-ablated surface~\cite{hua2020convex,zhao2019formation,S_LF_P25_Dudutis_2020,S_LF_P47_Zhao_2019}. The \ce{CO2} laser achieves this by selectively heating the surface of fused silica to a temperature above its softening point but below its melting; also, sublimation is not substantial. 
The 10.5~$\mu$m wavelength of \ce{CO2} laser is directly absorbed into Si-O vibrational band leading to a localised heating and surface reflow. 
Exposure conditions of \ce{CO2} were optimised for the reduction of roughness without change of lens shape as described in Sec.~\ref{Methods}.
This controlled reflow process significantly $\sim 50^\times$ reduced surface roughness to $R_a\sim 5$~nm, while maintaining the structural integrity of the lens as shown in Fig.~\ref{f-CO2}.

It is noteworthy, that similar surface smoothing can be made using silica coating of $h = 100-200$~nm thickness (comparable with roughness) over laser machined crystalline surface, e.g., \ce{Al2O3}. In such a case the width of laser scanned line $2b$ is important due to possible buildup of compressive stress and delamination of the coated film. The critical stress $\sigma_c = \frac{\pi^2}{12}\frac{E}{1-\nu^2}\left(\frac{h}{b}\right)^2$, where $E$ is the Young modulus of the coating (silica), $\nu$ is the Poisson ratio~\cite{MOON20043151}. A larger \ce{CO2} laser beam is preferable and the protocol of controlled melting and reflow needs optimisation, which is a simple procedure.   

\begin{figure*}[tb]
    \centering\includegraphics[width=12.5cm]{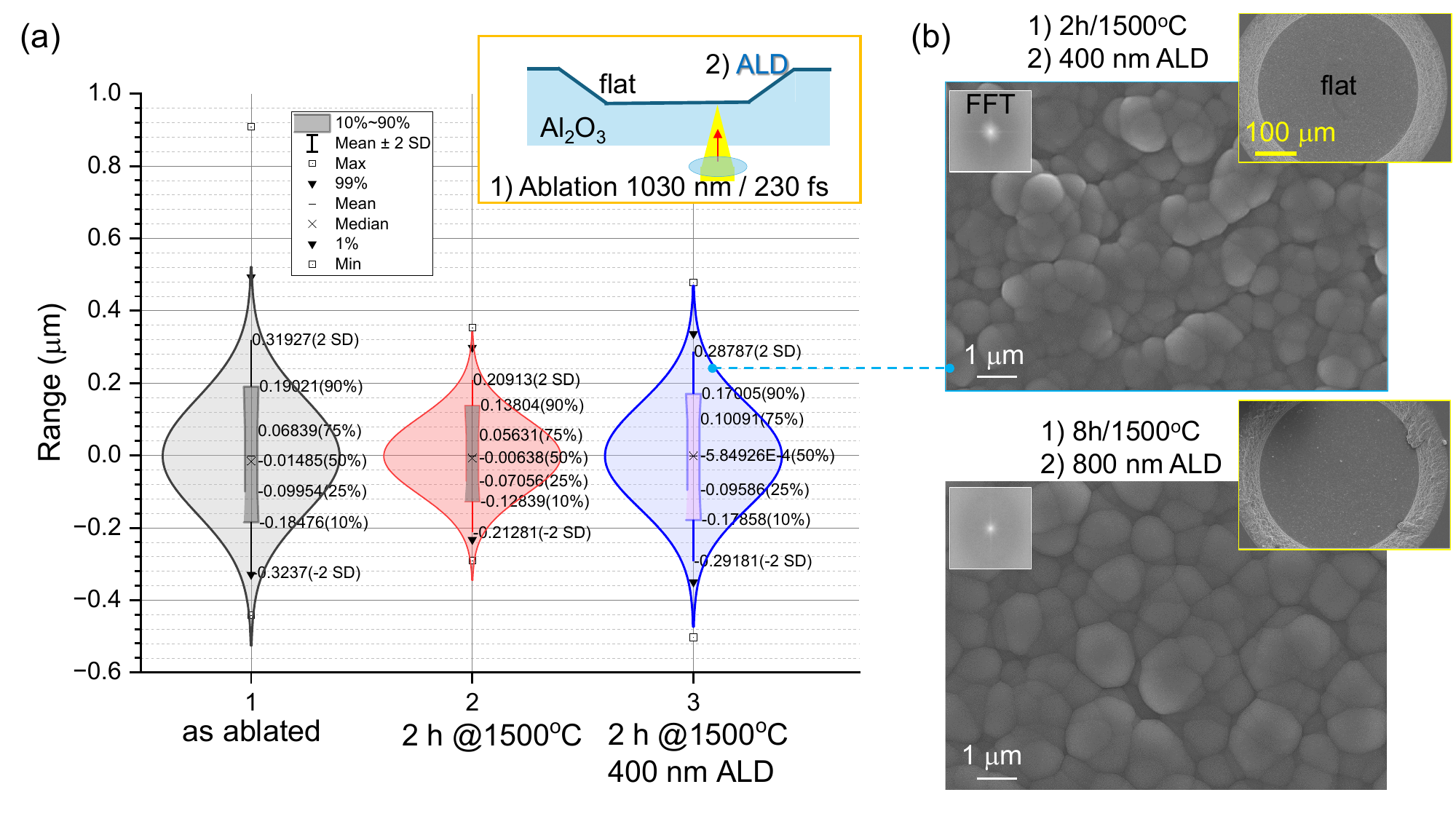} \caption{\label{f-hta} The high-temperature annealing (HTA) and atomic layer deposition (ALD). (a) Statistical analysis of surface roughness: as ablated, HTA, and ALD coated. Inset shows geometry of sample. (b) SEM images of ALD coated surfaces with 400~nm and 800~nm thickness. Insets show larger area of flat surface (pedestal of lens) and Fourier fast transform (FFT) of image. Optical profilimetry was carried out on surfaces coated by 10~nm of Cr. 
}
\end{figure*}
\begin{figure*}[tb]
    \centering\includegraphics[width=15.5cm]{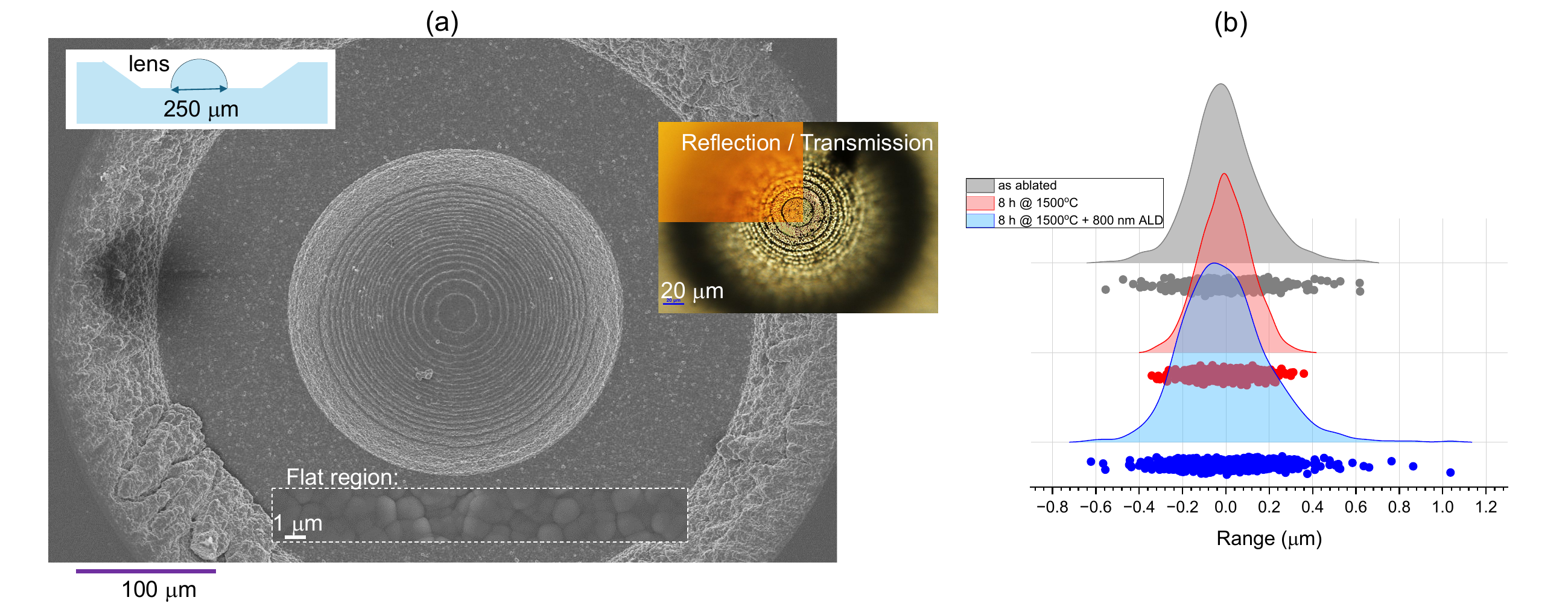}      \caption{\label{f-sap} Back-side ablated sapphire $\mu$-lens after HTA and ALD. Inset shows optical images in refection and transmission (a) SEM image of $\mu$-lens after 8~h/1500$^\circ$C and 800~nm alumina ALD coating. (b) Statistical analysis of roughness evolution by the normal distribution (made by toolbox in OriginPro 2023): as ablated, HTA, and after 800~nm ALD of alumina. The same binning experimental data was used (represented by dots-pattern).  
}
\end{figure*}

\subsection{High temperature annealing}

The high temperature annealing (HTA) is one of the methods for reduction of surface roughness. For crystals, it can be used to restore atomically flat surface and remove the smallest debris $< 20$~nm  by their re-incorporation and re-crystallisation into the atomic terraces of the ablated crater and, probably, partly by sublimation~\cite{23s2207968}. For sapphire, 1500$^\circ$C was used to remove debris after laser ablation when \ce{O2} flow is used~\cite{23p594}; it is noteworthy, that \ce{N2} atmosphere during HTA caused roughening of the surfaces and formation of \ce{AlON}. An additional benefit of such HTA treatment is the removal of blind spots for subsequent evaporation/sputtering of coating, including silica, which can be \ce{CO2} or HTA (at $\sim 1000^\circ$C) re-flown for smoothing the surface of a crystalline lens. Noteworthy, such HTA for silica reflow will have no influence on shape compromise of the crystalline surface. This virtue can find applications in crystalline optical elements.

\begin{figure*}[tb]
    \centering\includegraphics[width=16.5cm]{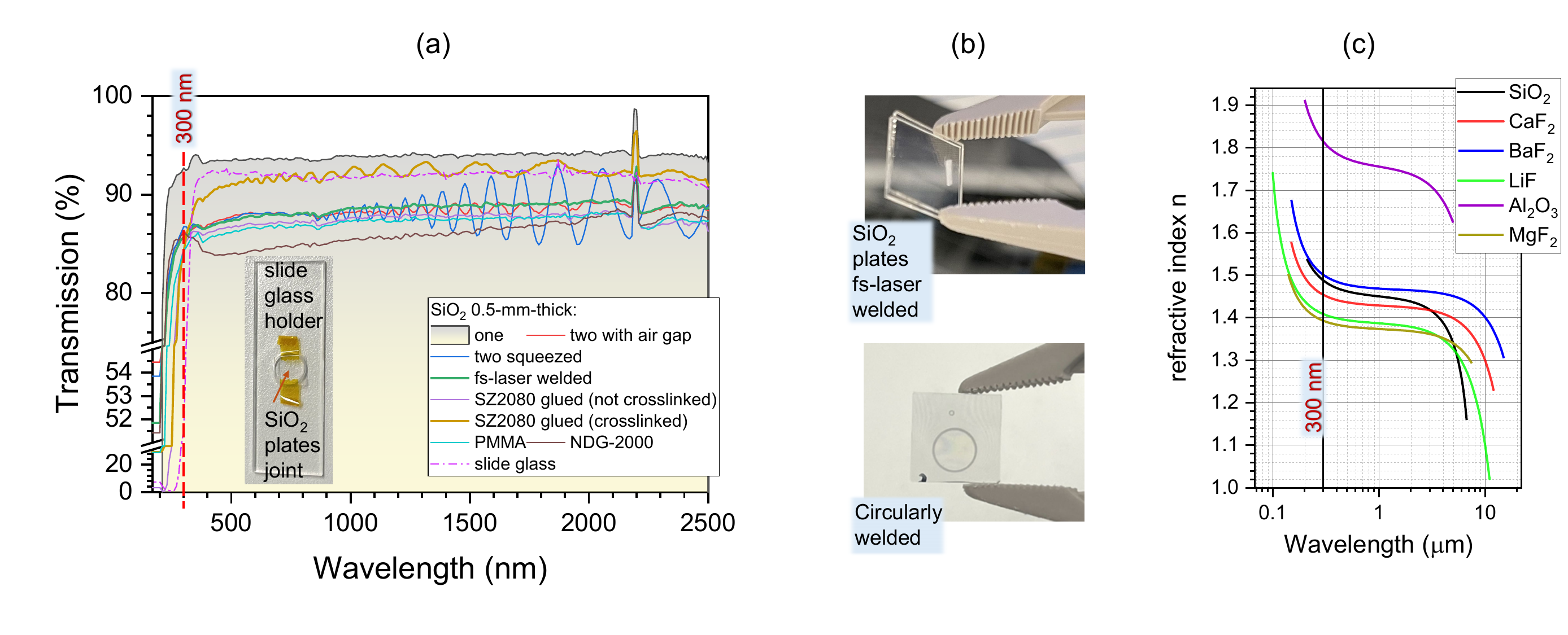}      \caption{\label{f-weld}  
(a) Splicing two silica slabs ($1\times 1$~cm$^2$) of 0.5~mm thickness by different methods: glue together (with PMMA resist, SZ2080\texttrademark resist without photoinitiators, spin-on glass NDG-2000) and fs-laser welding/joining. Samples were placed on a slide-glass holder with central hole and transmission was measured through joint section (or at close proximity for laser welded pieces). Crosslinking of SZ2080 was on a hot plat at $200^\circ$C overnight
. (b) Two 0.5~mm silica slabs welded (Conditions: 1030~nm/200~fs pulses, at repetition rate if $1$~MHz, pulse density of $200$~pulses/$\mu$m, burst per irradiation site was $3$, the average power of $10$~W (single-pulse fluence $9.7$~J/cm$^2$), at scanning speed of $0.5$~mm/s and line spacing $5~\mu$m. (c) Refractive indices $n$ of fluoride, silica and sapphire over UV-IR spectral range~\cite{nfo}. 
}
\end{figure*}
\begin{figure}[tb]
    \centering\includegraphics[width=8.5cm]{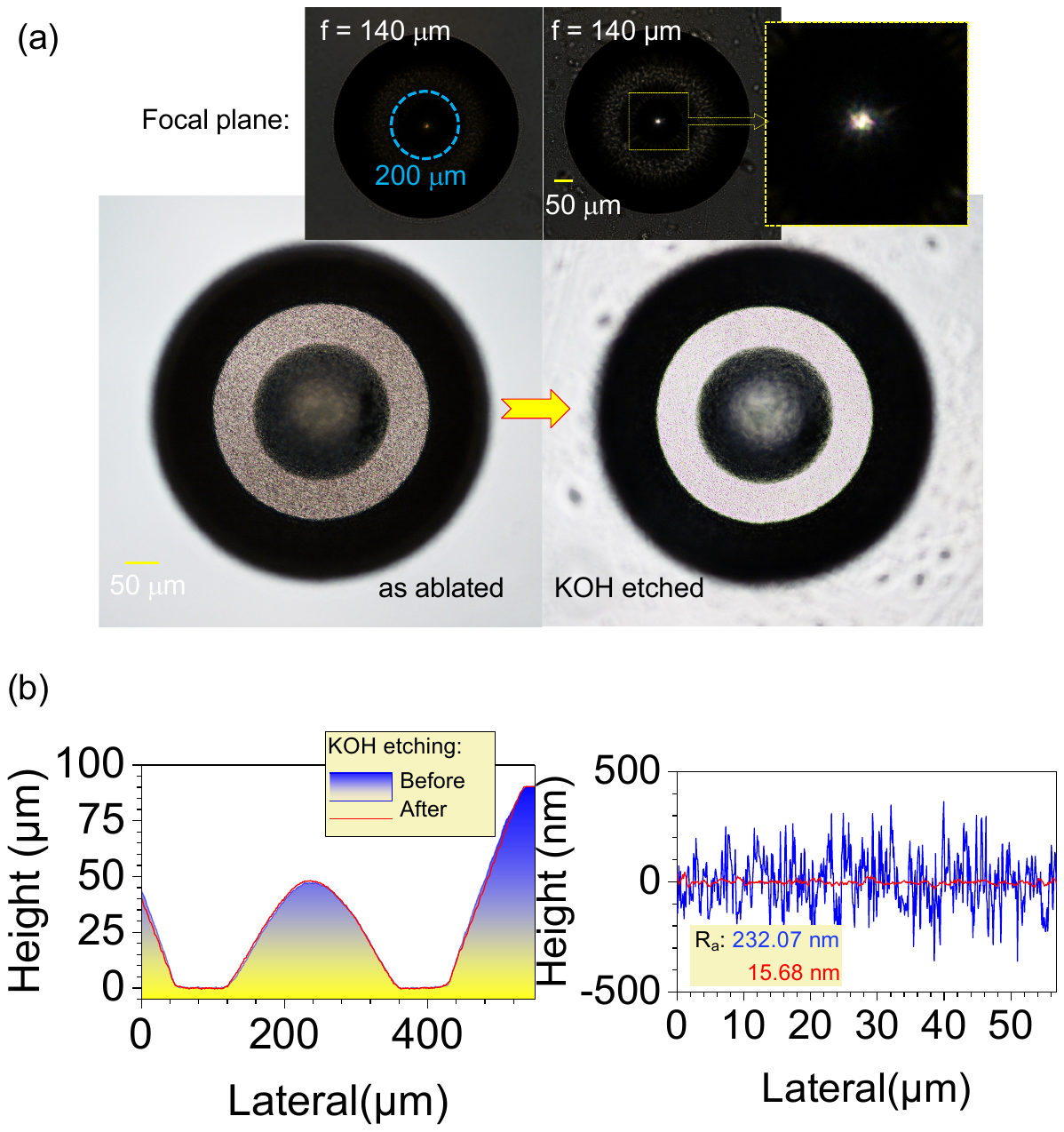}      \caption{\label{f-KOH} Back-side ablated sapphire $\mu$-lens before and after wet etch ($\sim 10$~min in 30\%wt. KOH in IPA - isopropyl alcohol).  (a) Optical image of the same lens before and after wet etch in KOH. 
    Top-insets show images of the focal plane. The close-up view has a cross-section approximately of $200-\mu$m-diameter fibre core. (b) Cross sections after wet etch smoothing for the lens and flat region (as in Fig.~\ref{f-Ra}). Roughness (rms) $R_a\approx 15.68$~nm was achieved without detrimental shape morphing of the spherical micro-dome.   
}
\end{figure}

The rationale behind surface reflow by silica on nano-rough surface of crystalline \ce{Al2O3} is based on arguments of surface tension (for liquid) or surface energy (solid), i.e., surface tension is force $F$ per length $l$ defining molecular interaction: $\gamma\equiv\frac{1}{2}\frac{F}{l}$~[N/m]. The larger the melting temperature of the solid, the larger surface energy, hence, sapphire is expected to be wetted by molten silica (lower melting temperature). The surface energy is linked to the wetting of surface via the Young's equation: $\sigma_s = \sigma_{sl}+\sigma_l\cos\theta_c$, where $\sigma_{s,l}$ is the surface energy of the solid and liquid (surface tension), respectively, $\sigma_{sl}$ is the interfacial tension between liquid and solid, and $\theta_c$ is the contact angle of the liquid on the solid. For non-polar materials, the Zisman method is used to determine surface energy~\cite{Zisman}. It assumes that the surface energy of a solid is equal to the maximum surface tension of a liquid at $\theta_c = 0^\circ$ (full wetting $\cos\theta_c\rightarrow 1$). This is known as the critical surface tension $\sigma_c$. Surface energy of \ce{Al2O3} at $1500^\circ$C was determined using Zisman's method with melted metal droplets and found $\sigma_c = 1.05 $~N/m~\cite{Eberhart}. Typical optical glasses have a lower surface tension of $\sim 0.22\pm 0.02$~N/m at $1300^\circ$C and a positive $d\sigma/dT > 0$ temperature coefficients~\cite{Zisman}; for comparison, \ce{H2O} at room conditions has surface tension $\gamma = 0.072$~N/m. Silica glass has a surface tension of $0.29$~N/m at $1100-1300^\circ$C~\cite{Boyd}. Hence, a liquid molten silica should wet nano-rough \ce{Al2O3} at high temperatures and create minimised surface, hence, nano-smooth surface.

Laser ablated and HTA treated surface of \ce{Al2O3} has a nano-crevice/capillary structure which favours advancement of a wetting liquid in time. The Lucas–Washburn model of liquid front advancement in time can be applied for the surface wetting~\cite{LW} defining the position of liquid front in time $t$: $L_f=\sqrt{Kt}$, where $K = r\frac{\gamma}{\eta}\frac{\cos\theta_c}{2}$, where $r$ is the radius of capillary/crevice, $\gamma$ is the surface tension of liquid, $\eta$ is the viscosity of the liquid, $\theta_c$ is the contact angle. This expression reveals scaling: smaller crevices (smaller $r$) require longer time to be filled, larger surface tension $\gamma$ and smaller viscosity $\eta$ favours a faster movement of the wetting front.

Figure~\ref{f-hta} shows a summary of surface roughness of laser back-side ablated 0.5-mm-thick \ce{Al2O3} slab and subsequent HTA and ALD of alumina. Statistical analysis was carried out by the normal distribution fit $f(x,\mu,\sigma) = \frac{1}{\sigma\sqrt{2\pi}}e^{-\frac{(x-\mu)^2}{2\sigma^2}}$, where $\sigma$ is the standard deviation (SD), $\mu$ is the average, $x$ is coordinate along the scan. Figure~\ref{f-hta}(a) shows that 2~hours at 1500$^\circ$C reduced initial roughness $2\sigma\approx 319$~nm (2SD)  to 209~nm. Interestingly, a subsequent ALD of 400~nm of alumina increased roughness $2\sigma = 288$~nm. If the ALD layer was 800~nm, $2\sigma = 396$~nm; see SEM images in (b) reveal a larger grain structure for a thicker ALD film. ALD films had isotropic orientation as revealed by fast Fourier transform (FFT). Figure~\ref{f-sap}(a) shows SEM image of the $\mu$-lens after HTA and 800~nm ALD. The fit of experimental data to the normal distribution reveals intricate changes of the surface roughness shown in Fig.~\ref{f-sap}(b). Kurtosis is a statistical measure that describes the shape of a distribution's tails and skews its asymmetry in relation to the overall shape. An apparent narrowing of the distribution is recognisable after HTA (Fig.~\ref{f-sap}(b)) with reduction of skewness from 0.34 (as ablated) to -0.08 while kurtosis reduced from 4.34 to 2.97. However 800~nm of ALD alumina increased the skewness and kurtosis to 0.66 and 5.01, respectively, which are larger than those measures on the as ablated surface. Discussion of roughness changes of ALD coatings is presented in Sec.~\ref{disco}. To reduce the final roughness of \ce{Al2O3} $\mu$-lenses, it was important to use 2-4~hours HTA since a longer $\leq 8$~h HTA caused increased roughness and shape change of statistical distribution (Fig.~\ref{f-50h}); with 4~h HTA the minimum value of $\sigma \approx 150$~nm was obtained.   
\begin{figure*}[t!]
    \centering\includegraphics[width=17cm]{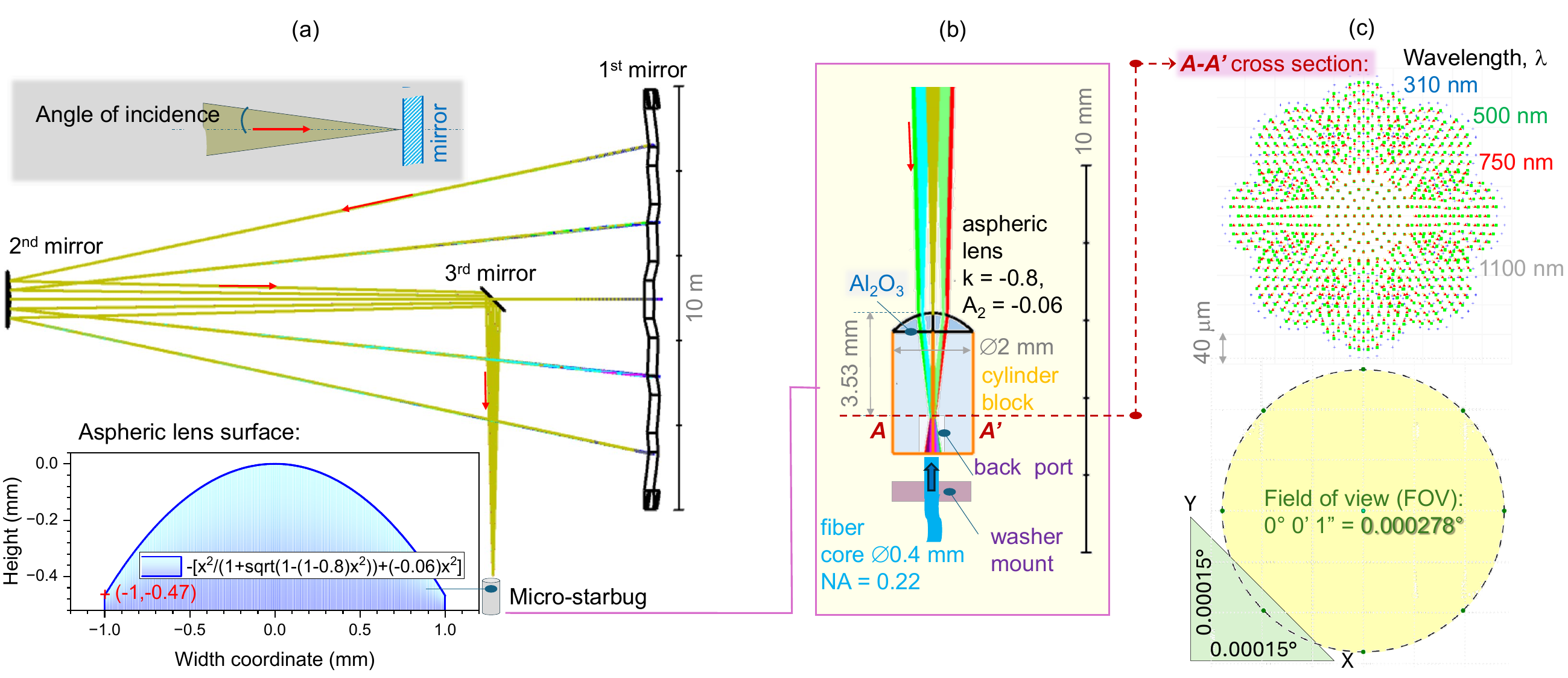}      \caption{\label{f-bug} Illustrative design for a $\sim$4-mm-long \ce{Al2O3} $\mu$-coupler for astronomy.   (a) ZEMAX simulation of the Keck telescope with a segmented 10.95 m diameter primary mirror composed of 36 hexagonal 1.8~m segments; optical design~\cite{CAS}.  Light is reflected from the primary to the secondary mirror and then to the Nasmyth focus 
    where the $\mu$-coupler is placed. The inset shows the aspheric surface on the 0.5-mm-thick \ce{Al2O3} front lens of the $\mu$-coupler. (b) ZEMAX simulation of the $\mu$-coupler, consisting of a $~\sim 6$~mm and 2~mm diameter sapphire tube with an aspherical lens laser ablated on the back-side of 0.5-mm-thick sapphire disk (inset in (a)). 
    The blue (310 nm), green (500 nm), red (750 nm), and yellow (1100 nm) light cones correspond to the wavelengths required for the FOBOS instrument. 
    (c) Full-field spot diagram showing how the 1 arcsecond field of view (FOV) is focused as it enters the 0.4~mm fibre; colour markers represent the corresponding wavelength. The green dots in graph underneath show the location of where the cones of light at the specific wavelengths will be coming from this is reflected in (a) where the 1 arcsecond field of view comes from the telescope's primary mirror.  
}
\end{figure*}

\subsection{Femtosecond laser welding/joining}

Splicing of two (or more) sections of 3D optical element is required when back-side ablation is used, since 0.5~mm is a practical thickness when laser ablation can be made with small focal spot 2-3~$\mu$m ($NA\leq 0.4$) needed for high precision fs-laser machining. Different  ``glues'' were tested: PMMA resist, SZ2080\texttrademark resist without photoinitiator and spin on glass NDG-2000. Results of transmission measurements over UV-IR spectral range are shown in Fig.~\ref{f-weld}. To have high $T>85\%$ at short wavelength $\lambda \approx 300$~nm (see marker in Fig.~\ref{f-weld}) is demanding task. Also, there is a gap between two segments which should be minimised and/or refractive index matching glue should be used to minimise Fabry-Perot fringing. The gap, of thickness $d_{gap}$ with refractive index $n_{gap}$, will cause interference. With two selected adjacent maxima (or minima) at $\lambda_{1,2}$, the thickness $d_{gap} = \frac{1}{2[1/\lambda_1 - 1/\lambda_2]n_{gap}}$~\cite{24sp059}. For the most prominent interference pattern with air gap in between \ce{SiO2} slabs (Fig.~\ref{f-weld}(a)), $d_{air} = 10.3~\mu$m  with $\lambda_{1,2}= 1870.8; 2058.9$~nm. For the cross-linked SZ2080 with $n_{SZ2080} \approx 1.5$ at visible range, $d_{SZ2080} = 2.7~\mu$m for $\lambda_{1,2} = 650.5; 706.9$~nm. These estimates show that fs-laser welding has to produce large structural changes at the surface in order to join two pieces of silica by mixing molten regions.

During the laser welding/joining of silica, a high pulse density of 
200 pulses per 1~$\mu$m scan length at 1~MHz repetition rate was shown to enhance 
bonding in fs-laser welding by delivering energy in rapid succession to a small area. This results in intense localised heating
which raises the temperature at the material interface, facilitating improved fusion and stronger bonding. With high pulse density, energy is concentrated over shorter time intervals, which reduces heat diffusion into surrounding areas. Consequently, this localised heating minimises thermal gradients and potential material damage, while promoting a more robust bond at the interface. 
Since fs-laser welding/joining is a direct write method, a circular trajectory is useful for splicing two parts of an optical element centred around the optical axis. 

\subsection{Sapphire \texorpdfstring{$\mu$}{TEXT}-lens}

Back-side ablation of \ce{Al2O3} produced surface with $\sim 230$~nm roughness when same laser fabrication conditions were used as for \ce{CaF2} and silica $\mu$-lenses (Fig.~\ref{f-KOH}). Wet etch of the ablated surface in 30\%wt. KOH solution of isopropyl alcohol (IPA) for $\sim 10$~min smoothed surface to $R_a\approx 15$~nm, which is an acceptable $\lambda/10$ surface roughness even for UV wavelengths. Remarkably, the focal length of the lens was not changed, what indicates that shape was not altered. Laser amorphised regions of \ce{Al2O3} can be etched in KOH solution~\cite{08pssrrl275}, which is preferred to high $245^\circ$C temperature etch in \ce{H2SO4}:\ce{H3PO4} (3:1)~\cite{ZHANG}.  

The design of an illustrative fibre $\mu$-coupler is shown in Fig.~\ref{f-bug}. The inset in (a) shows the aspherical profile of the front surface for a 0.5-mm-thick \ce{Al2O3} slab as used in this study for the back-side ablation. The optimisation was carried out in ZEMAX for $A_2$ parameter with higher orders zeroed and $k=0$ using height profile along the cross section: $y(x) = \frac{Cx^2}{1+\sqrt{1-(1+k)C^2x^2}} + A_{2}x^2$ for collecting all rays onto $0.4$-mm-diameter spot. It was found by numerical experimentation that higher orders of aspheric parameters were not required. 
The Keck ray-path model in ZEMAX accounts for the focal ratio degradation (FRD). FRD describes the spreading of light injected at a given focal ratio (or numerical aperture), causing it to emerge at a lower focal ratio. This broadens the light output, reduces throughput, impacts coupling efficiency, degrades image quality, and increases modal noise~\cite{Staszak_2016,Minardi_2021}. The exit angle of the device relative to the telescope focus is approximately 1.18$^\circ$, which is significantly lower than the critical value of 9.46$^\circ$. This ensures that light leakage due to FRD is negligible. 
For comparison with a spherical lens, an estimate is following for the focus placed at $f/R = 2.3$ (inside \ce{Al2O3}) for the central ray (Fig.~\ref{f-inside}(b)): $f = 4.6$~mm for $R=1$~mm $\mu$-coupler. The side rays are focused on $f/R = 1.7 \approx 3.4$~mm. The narrowest focal region is within the axial range of 4.6-3.4~mm.

For the aspheric lens, the surface is defined with $k = -0.8$ (prolate spheroid for $-1<k<0$) and $A_2 = -0.06$, which determines the focal cross-section (\emph{A-A'}) of 0.4~$\mu$m at the axial position of 3.53~mm (Fig.~\ref{f-bug}(b)). The final $\mu$-coupler can be made from a front lens with surface ablation using back-side irradiation through a 0.5-mm-thick slab spliced with a $\sim 3.5$~mm cylinder block, which has a back-side fibre port milled by direct laser ablation (a front-side ablation). The back-port is designed to be aligned onto the optical axis and makes a simple assembly with an optical fibre. A washer (plastic) can be added for the mechanical strength of all assembly. The optical resist SZ2080\texttrademark~ will be used for \ce{Al2O3} block connection to the fibre with a thickness of a few micrometers (tested in welding; Fig.~\ref{f-weld}). As expected, the optimised focal position for the aspheric surface is within the range predicted for the simplest spherical lens under plane wave illumination (Fig.~\ref{f-inside}). The caustic region for 1" incident beamlets coloured according to their wavelengths 310-1100~nm were all collected onto the aimed 0.4-mm-diameter fibre core (Fig.~\ref{f-bug}(c)). 

It was also possible to collect all rays onto $\sim 120~\mu$m diameter at $\sim 3$~mm from tip of lens using $k = -0.85$, $A_2 = -0.13312$, which corresponded to the depth of $y = -0.38718$~mm at $x=\pm 1$~mm (not shown here for brevity). For back-side ablation, it was important that the required surface profile is not exceeding 0.5-mm-thickness, which is practical for comparatively high-$NA$ ablation (top-inset in Fig.~\ref{f-bug}(a)). The angular values from the ZEMAX model for all optical pathways, including the aspheric lens with focus inside the sapphire tube 
are the angles of incidence (top-inset in Fig.~\ref{f-bug}(a)). 

\section{Discussion}\label{disco}
\subsection{ALD coatings}

ALD is known to coat complex 3D surfaces conformally and is used for thin few nanometres modifications of nanostructured elements and hundreds-of-nm for $\mu$-elements as multi-level microlenses~\cite{Astrauskyte}. However, 
thick \SI{2}{\micro\metre} 
ALD coatings were also used to produce multilayer structures over initially corrugated nanostructured patterns. It was observed that after a 500~nm thickness layer, the initial modulation was almost flattened~\cite{Lina}. This was the motivation for depositing the ALD alumina layer on sapphire surface, which are primarily smoothed by HTA, for void-free smoothing coatings. However, the roughness was not reduced and even increased with thick 800~nm ALD coatings. This observation calls for deeper investigation, which is beyond this study. One possible explanation hints at the shape change of the normal fits of HTA treated surfaces when re-crystallisation occurred (Fig.~\ref{f-50h}). Change of symmetry of the normal distribution fit favours positive (bulging out) x-values, rather than filling deep pits. This is due to the growth of nano-/micro-pillars as can also be recognised in the SEM images of 50~h HTA treated $\mu$-lenses. Formation of step-terraces by HTA might influence formation of alumina ALD coating over crystalline \ce{Al2O3} surface. Since sapphire tends to crystallise, the formation of disordered alumina was less probable and crystalline-like growth was favoured. This caused increase in surface roughness. Use of dissimilar materials between substrate and ALD coating could clarify this issue.    

\subsection{Down-sized \texorpdfstring{$\mu$}{TEXT}-lens fibre couplers}

Finally, with established $k,A_2$ parameters for the aspheric lens using ZEMAX model for the Keck optical path (Fig.~\ref{f-bug}), the wave based methods were used to explore the formation of the focal region for the case study $k = -0.85$, $A_2 = -0.13312$. This is important for miniaturised $\mu$-optical devices and cannot be modelled with ray-optics methods and software such as ZEMAX, OSLO, etc. Indeed, the focal length is determined only by the radius of the spherical focusing surface and refractive index of a lens in ray-optics definition $1/F = (1-n)/R$. The finite difference time domain (FDTD; Lumerical, Ansys) was used to explore the wave nature of focal region formation at the extreme case when the aspheric shape with the same coefficients is down-sized to a few wavelengths scale (by a factor $10^3$ from the actual $\mu$-coupler, hence, millimetres become micrometres for the lens) shown in Fig.~\ref{f-fdtd}. Localisation of the focal region at $\sim 1.7~\mu$m along propagation inside sapphire for the $R = 1~\mu$m radius lens (would correspond to $\sim$1.7~mm for $\mu$-coupler of $R = 1$~mm) with width well fitting into a 0.4~$\mu$m (actual 0.4~mm) fibre core is evident for targeted UV 310~nm wavelength for the incident plane wave. This modelling is valid for the Fresnel diffraction near-field regime when Fresnel number $F = \frac{a^2}{L\lambda}\ll 1$, where $a$ is the aperture (lens) size, $L$ is the distance to the screen (fibre entrance) for wavelength $\lambda$ (see simulations using Fresnel integral (Sec.~\ref{Fresn}) in Fig.~\ref{f-Fres}). Another method to explore light propagation is via the Rayleigh-Sommerfeld integral, which can predict focusing for $\mu$-optics~\cite{24m798}, however, was not suitable for the 2-mm-diameter lens where the ray-optical model is adequate.   

\section{Conclusions and outlook}

A toolbox of different techniques is established for fs-laser back-side ablation in air or liquid of 3D free-form $\mu$-optical elements out of different materials glasses and crystals for UV-IR spectral window. They include HTA, ALD coating, laser morphing/smoothing,  magnetorheological polishing, and wet etching. Different parts of an entire optical element can be assembled by fs-laser welding/joining. This is especially important when back-side laser machining with high precision is made since this requires the thickness of a sample slab less than 0.5~mm for most of practical applications. Also, cw-laser \ce{CO2} can be used for surface smoothing of silica down to roughness of a few nanometres. All processes are practical and allow to make 3D $\mu$-optics with a $\lambda/10-\lambda/20$ surface finish at the required wavelength. As a test case, we aimed at prototyping processing steps for a $\mu$-coupler optical fibre coupled element for the planned UV ($\sim 300$~nm) FOBOS spectrometer application. 

The front-lens of the $\mu$-coupler has $NA = n\sin\alpha$ matching the numerical aperture $NA = 0.22$ of the optical fibre with $0.4~\mu$m diameter core and corresponds to the half-angle of light collection cone $\alpha\approx 10^\circ$. Back-side laser machining was carried out on 0.3-0.5 mm slabs of glass and crystal. The process was performed in air at single pulse fluences in the range of $15-25$~J/cm$^2$ (average intensity $70-130$~TW/cm$^2$) with a high pulse overlap of $\sim 89.3\%$. A systematic study of surface roughness and its reduction to $\lambda/10 - \lambda/20$ range via \ce{CO2} laser melting, magnetorheological polishing, thin film evaporation/sputtering, and high temperature annealing, and wet etching shows strategies to tailor surfaces for the final UV-IR application. Splicing of different materials for $\mu$-coupler via fs-laser welding/joining is demonstrated. All toolbox of processing steps with fs-laser machining defining the required geometry of 3D free-form element is developed. 

Multi-lens array is a possible scale up of the demonstrated method. At the optimised fs-laser machining conditions, it took 15~min for $0.25$~mm diameter $\mu$-lens (on any substrate) and would require $\sim 16$~hours for the final 2-mm-diameter $\mu$-coupler. A higher repetition rate and proportionally faster scanning could be used for the developed protocol (fixed ablation dose). It should take less than $\sim$5~hours for practical application. Such conditions are available for typical modern fs-fab setups. Also, scanning protocols can be further improved for a specific method used for final surface smoothing.       
 
\small\begin{acknowledgments}
This study was carried out via ARC DP240103231 grant. This work was performed on fs-fab laser station at Swinburne which is part of the Melbourne Centre for Nanofabrication (MCN) in the Victorian Node of the Australian National Fabrication Facility (ANFF). 
We thank Workshop-of-Photonics (WOP) Ltd., Lithuania for the patent licence and technology transfer project by which the industrial fs-laser fabrication setup was acquired for Nanolab, Swinburne. We acknowledge discussions of Starbug performance with Prof. Kevin Bundy. 
\end{acknowledgments}

\appendix
\setcounter{figure}{0}
\makeatletter 
\renewcommand{\thefigure}{A\arabic{figure}}
\begin{figure*}[h!]
    \centering\includegraphics[width=18cm]{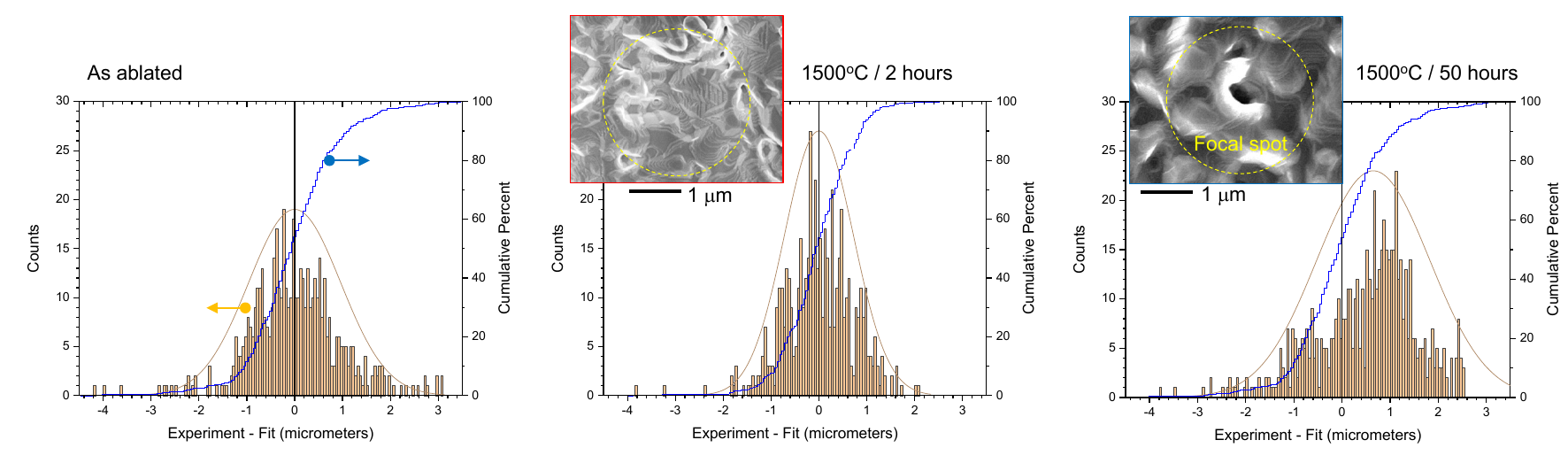}      \caption{\label{f-50h} Statistical analysis of surface roughness on the back-side ablated sapphire 250-$\mu$-diameter lens after 1500$^\circ$C HTA for 2~h and 50~h; HTA was carried out in tube furnace (Nabertherm) in \ce{O2} flow at 20~mL/min. The normal distribution fit $f(x,\mu,\sigma) = \frac{1}{\sigma\sqrt{2\pi}}e^{-\frac{(x-\mu)^2}{2\sigma^2}}$, where $\sigma$ is the standard deviation (SD), $\mu$ is the average, $x$ is coordinate along the scan. Inset shows SEM images after HTA from the top of the lens; the size of a focal spot of laser machining  $1.22\lambda/NA\approx 2.8~\mu$m is shown. Fabrication: 1030~nm/230~fs, $NA = 0.45$, write speed 10~mm/s, laser repetition rate 200~kHz, pulse 
    fluence 24.5 J/cm$^{2}$ on the sample per pulse, 3 pulses per ablation site. Milling conditions: step height between neighbouring lines 0.7~$\mu$m, number of passes on the same height 1, lateral shift between neighbouring lines 0.7~$\mu$m. The normal distribution $\sigma = 0.96~\mu$m (as ablated), 0.72~$\mu$m (2~h HTA), 1.17~$\mu$m (50~h HTA). A step in height for $0.9~\mu$m during laser ablation caused $\sim 1.5~\mu$m step in \ce{Al2O3} due to refractive index $n\sim 1.7$.     
}
\end{figure*}
\begin{figure*}[tb]
    \centering\includegraphics[width=15cm]{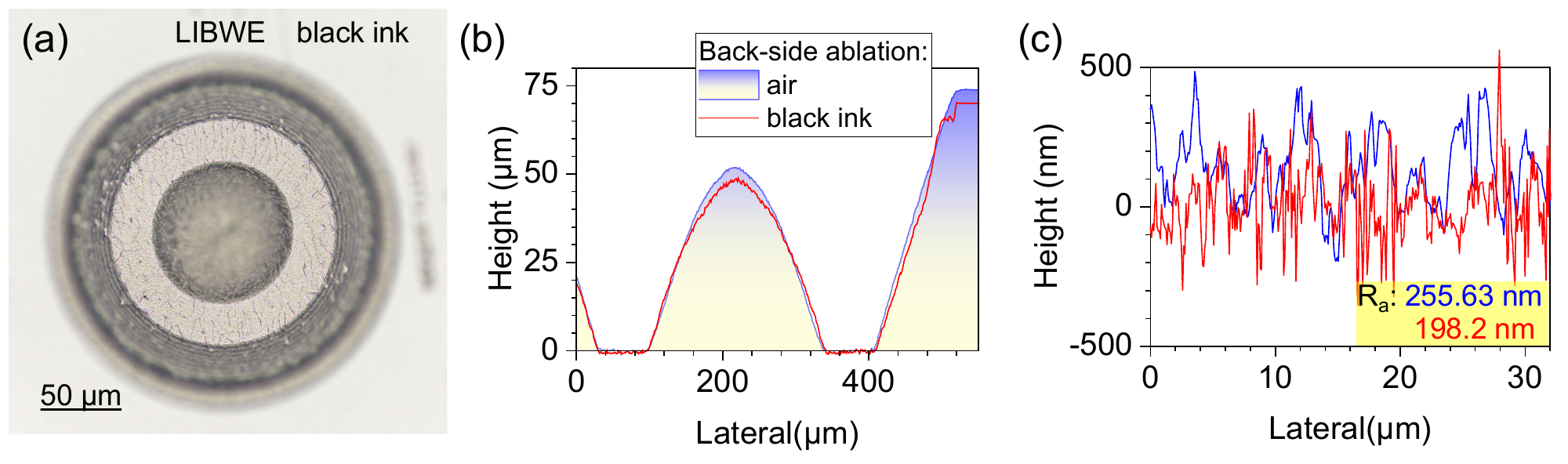}      \caption{\label{f-ink} Laser-induced back-side wet etching (LIBWE) with carbon black (sumi) and acetone solution mixed in a 1:1 ratio;  the offset parameter $Z_{offset} = 25~\mu$m. 
    (a) Optical image of the silica lens. Cross sections of the lens (b) and flat region (c) (as in Fig.~\ref{f-Ra}). Roughness (rms) $R_a\approx 198$~nm, achieved by LIBWE while $R_a\approx 256$~nm for back-side ablation in air.
}
\end{figure*}
\section{Long HTA}

Long 50~h HTA at 1500$^\circ$C was carried out. Evolution of surface roughness over the same trajectory (averaged over a 3 pixel width) of central cross section over the spherical lens was carried out: 1) as ablated, 2) HTA for 2~h, and 3) HTA for 50~h in \ce{O2} flow (Fig.~\ref{f-50h}). After 2~h HTA a narrower normal distribution function was fitting the experimental data. However, after 50~h, surface morphology become more coarse in terms of surface nano-relief structures. Apparent growth presumably along the fast direction $\left<111\right>$ was observed (see SEM insets in Fig.~\ref{f-50h}).      

\begin{figure*}[tb]
    \centering\includegraphics[width=13cm]{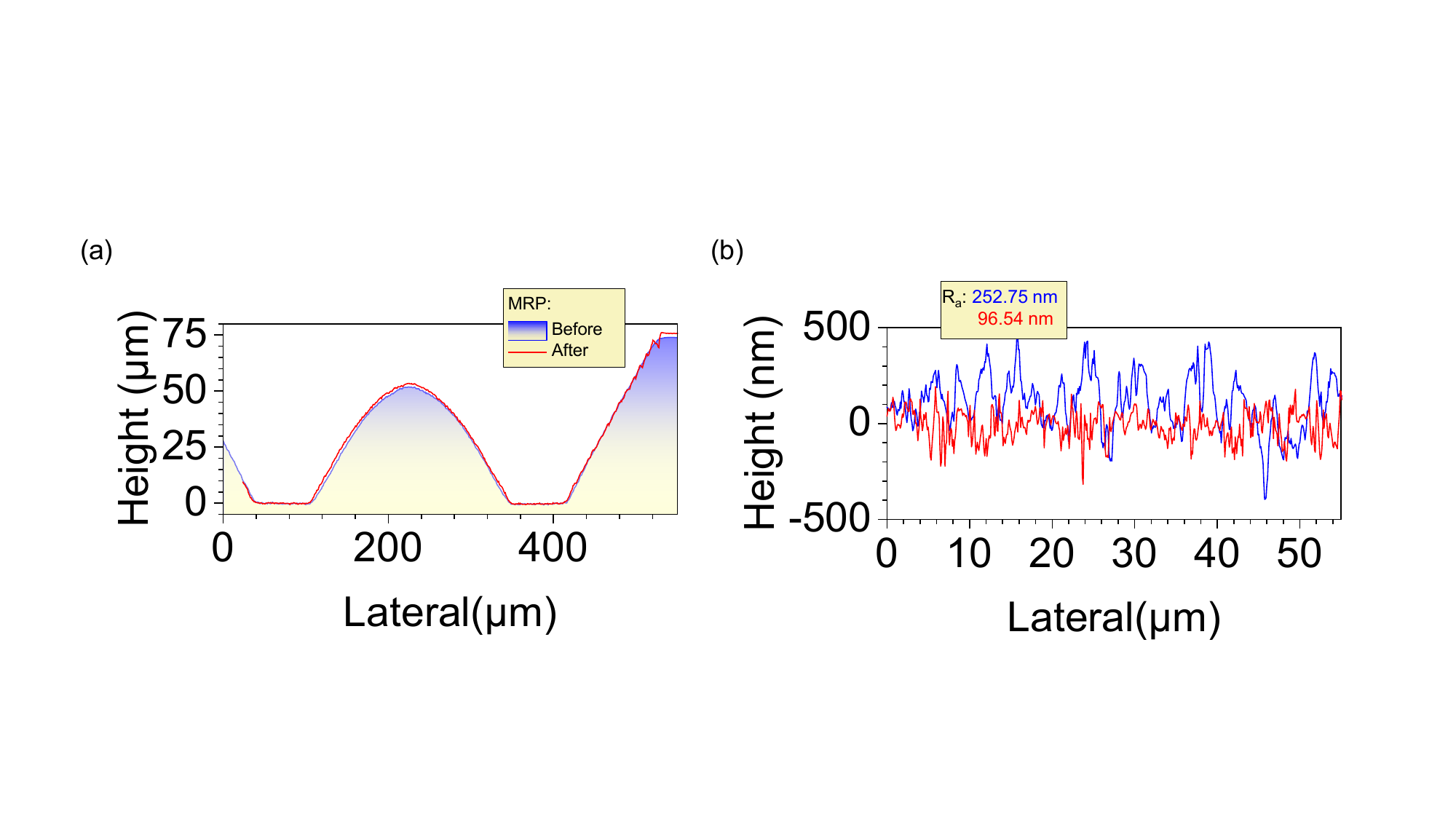}      \caption{\label{f-magn} Magnetorheological polishing (MRP).
    Cross sections of after MRP for the lens (a) and flat region (b) (as in Fig.~\ref{f-Ra}). Roughness (rms) $R_a\approx 96.5$~nm was achieved without detrimental shape morphing of the spherical micro-dome.
}
\end{figure*}
\begin{figure*}[tb]
    \centering\includegraphics[width=17cm]{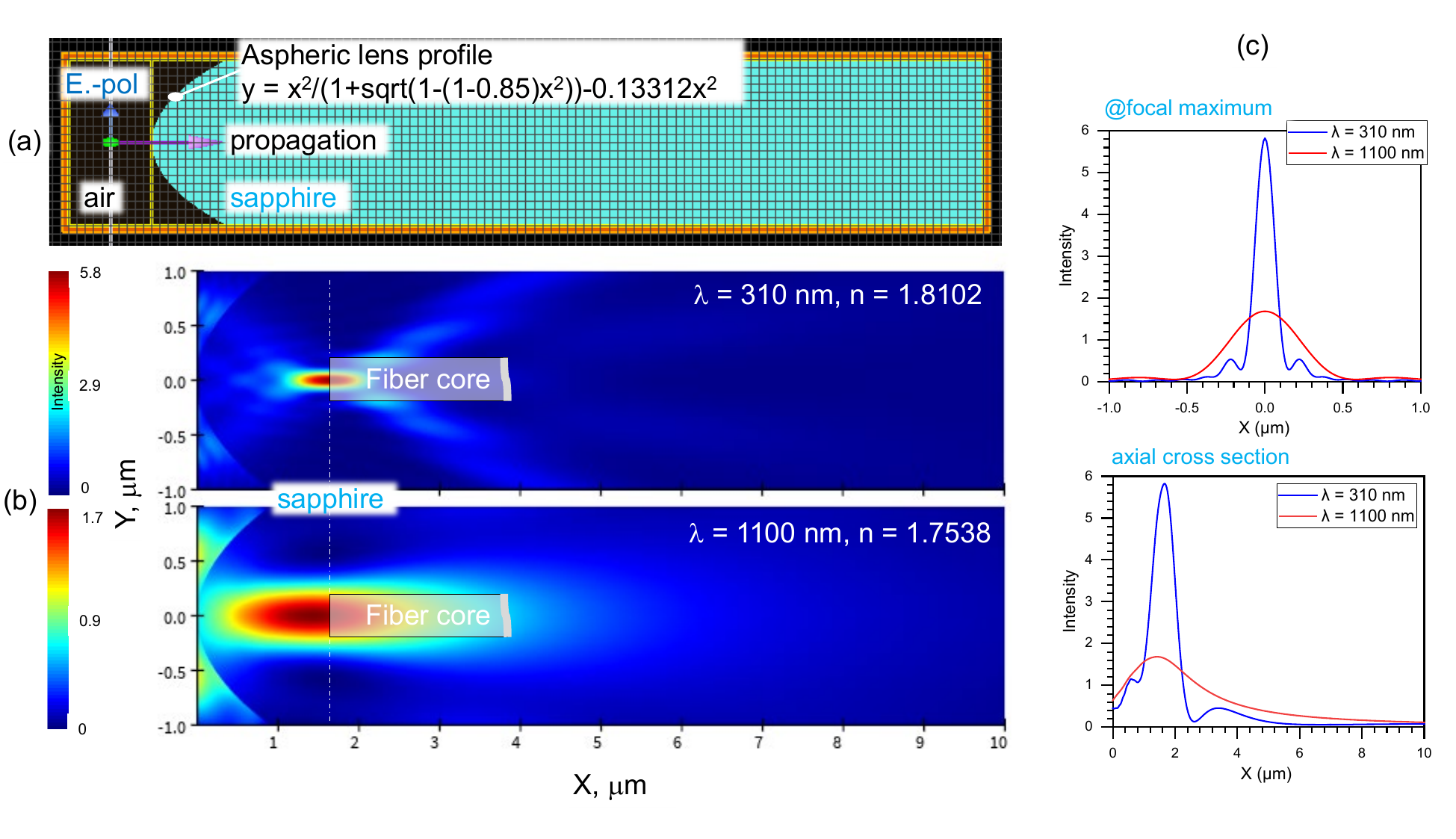}      \caption{\label{f-fdtd} A toy-model of down-sized aspheric lens by a factor of $10^3$ (mm becomes $\mu$m) with the same $k = -0.85$ and $A_2= - 0.1331$ coefficients and $C = 1~\mu$m$^{-1}$ as for the actual 1-mm-radius coupler lens. (a) Geometry of FDTD calculations (Lumerical, Ansys) for the plane wave incidence with electric field amplitude $E = 1$. (b) Calculated intensity cross sections for two wavelengths 310~nm and 1100~nm. The profile of a fibre core (dow sized to 0.4~$\mu$m from 0.4~mm in actual $\mu$-coupler) is overlayed for visualisation of overlap with the focal region. (c) Lateral and axial cross sections of intensity profiles at their maxima.}
\end{figure*}
\begin{figure*}[tb]
    \centering\includegraphics[width=17cm]{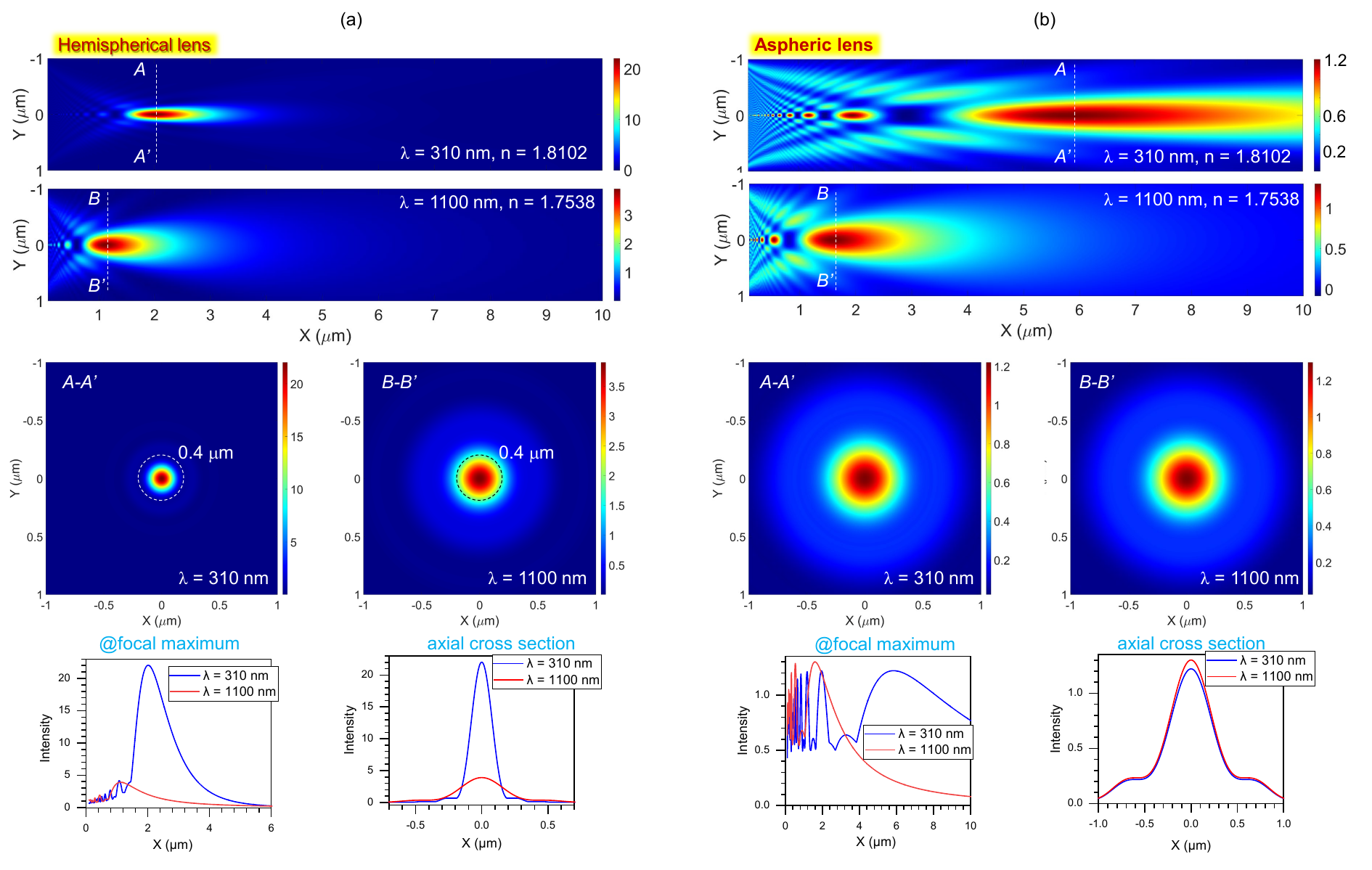}      \caption{\label{f-Fres} Fresnel diffraction of a down-sized hemispherical (a) and aspheric (b) lenses by a factor of $10^3$ (mm becomes $\mu$m). For aspheric lens, the parameters are $k = -0.85$ and $A_2= - 0.1331$ and $C = 1~\mu$m$^{-1}$ as for the actual 1-mm-radius coupling lens. The axial and lateral cross sections of the intensity profiles and lateral cross sections at the maxima for two wavelength. The dashed-circle marker in (a) shows relative diameter of the fibre core (actual 0.4~mm becomes 0.4~$\mu$m in the model). }
\end{figure*}

\section{Fresnel integral}\label{Fresn}

The Fresnel integral nomenclature is introduced next. Near-field light intensity distribution is simulated at the plane wave incidence for the $10^3$ times down sized $\mu$-lenses. 

The Fresnel integral relates the incident field to the diffracted field at any point along the propagation direction. For a lens, this involves computing how the incident light is altered by the lens's aperture and how it converges to form a focused spot at the focal plane. When light passes through the lens aperture, it experiences diffraction, which is characterised by the wavefront curvature. The diffraction integral accounts for how the wave converges, and its intensity distribution at the focal plane is calculated by integrating the field at the lens aperture.

When a plane wave of amplitude $U_0$ = 1 is incident upon a lens, two physical changes occur to the light field that impinges on the lens. The first change is the phase change of the field due to the change in the optical path, and the second is the amplitude change of the field due to the Fresnel reflection and transmission on the surface of the lens. The transmittance of a lens can be expressed as a complex function $t(x,y)$:
\begin{equation}\label{eqn_transmittance}
    t(x,y) = \frac{U_1(x,y)}{U_0(x,y)},
\end{equation}
where $U_0(x,y)$ and $U_1(x,y)$ are the light fields in the planes immediately before and behind the lens, respectively. In particular, we can express $t(x,y)$ as
\begin{equation}\label{eqn_modulations}
    t(x,y) = P(x,y) e^{-i\phi(x,y)},
\end{equation}
here $i = \sqrt{-1}$ is the imaginary unit, $P(x,y)$ and $\phi(x,y)$ are the two functions responsible for the amplitude and phase changes in the incident light, respectively. The function $P(x,y)$ is sometimes called the lens pupil function and is confined to the aperture of a lens. For a circularly symmetric lens, the pupil function is only a function of the radial coordinate ($P(x,y) = P(r)$), where $r = \sqrt{x^2 + y^2}$.
Since sapphire is highly transparent in the UV to IR regions, its absorption coefficient $\alpha$ is very small (close to 0) and we can ignore absorption and amplitude modulation $P(r) = 1$ assumed (no amplitude changes).
Therefore, the light field on the diffraction plane (immediately behind the lens):
\begin{equation}\label{eqn_U_1}
U_1 (r_1)= e^{-i\phi(r_1)} \cdot U_0(r_0),
\end{equation}
here $U_0(r_0)$ = 1, for a plane wave incident and $\phi(r_1)$ is the phase modulation:

\[
\phi(r_1) =
\begin{cases} 
n_{sapphire}\cdot k\cdot \left(f - \sqrt{r_1^2+f^2}\right),  & \text{(for hemispheric lens)}, \\
n_{sapphire}\cdot k\cdot\left[\frac{Cr_1^2}{1+\sqrt{1-(1+k)Cr_1^2}} + A_2r_1 ^2\right], & \text{(for aspheric lens)},
\end{cases}
\]
where $k = \frac{2\pi}{\lambda_{eff}}$ is angular wavenumber (spatial frequency), $\lambda_{eff} = \frac{\lambda}{n_{sapphire}}$ is the effective wavelength (for focusing inside sapphire), $f$ is the focal length of the hemispherical lens.

The light field $U_2 (r_2)$ on the observation plane (focal plane):
\begin{equation}\label{eqn_Fresnel}
U_2 (r_2)= \frac{i2\pi}{\lambda_{eff} z}\cdot e^{-ikz}\cdot e^{-\frac{ikr_2^2}{2z}}  \int_0^\infty U_1 (r_1)\cdot  e^{-\frac{ikr_1^2}{2z}}\cdot  J_0\left(\frac{kr_1r_2}{z}\right) r_1dr_1,
\end{equation}
where $z$ is the distance between the diffraction plane and the observation plane (equal to the focal length $f$ for focusing), $J_0$ is a Bessel function of the first kind of order zero. 

As a result, the intensity distribution $I$ on the focal plane that is the modulus square of $U_2 (r_2)$:
\begin{equation}\label{eqn_Intensity}
I = |U_2 (r_2)|^2.
\end{equation}
Figure~\ref{f-Fres} shows summary of Fresnel integral calculated for hemisphere and aspheric lenses at two different wavelengths (own Matlab code).

The Rayleigh--Sommerfeld (RS) diffraction integral~\cite{gu2000advanced} can be used to simulate the focal intensity distribution and, compared with Fresnel diffraction integral, the RS diffraction theory provides more accurate light diffraction predictions, because it does not assume the paraxial approximation~\cite{wei2021high}.

\clearpage
\bibliography{bib,paper6}

\begin{thebibliography}{48}%
\makeatletter
\providecommand \@ifxundefined [1]{%
 \@ifx{#1\undefined}
}%
\providecommand \@ifnum [1]{%
 \ifnum #1\expandafter \@firstoftwo
 \else \expandafter \@secondoftwo
 \fi
}%
\providecommand \@ifx [1]{%
 \ifx #1\expandafter \@firstoftwo
 \else \expandafter \@secondoftwo
 \fi
}%
\providecommand \natexlab [1]{#1}%
\providecommand \enquote  [1]{``#1''}%
\providecommand \bibnamefont  [1]{#1}%
\providecommand \bibfnamefont [1]{#1}%
\providecommand \citenamefont [1]{#1}%
\providecommand \href@noop [0]{\@secondoftwo}%
\providecommand \href [0]{\begingroup \@sanitize@url \@href}%
\providecommand \@href[1]{\@@startlink{#1}\@@href}%
\providecommand \@@href[1]{\endgroup#1\@@endlink}%
\providecommand \@sanitize@url [0]{\catcode `\\12\catcode `\$12\catcode `\&12\catcode `\#12\catcode `\^12\catcode `\_12\catcode `\%12\relax}%
\providecommand \@@startlink[1]{}%
\providecommand \@@endlink[0]{}%
\providecommand \url  [0]{\begingroup\@sanitize@url \@url }%
\providecommand \@url [1]{\endgroup\@href {#1}{\urlprefix }}%
\providecommand \urlprefix  [0]{URL }%
\providecommand \Eprint [0]{\href }%
\providecommand \doibase [0]{http://dx.doi.org/}%
\providecommand \selectlanguage [0]{\@gobble}%
\providecommand \bibinfo  [0]{\@secondoftwo}%
\providecommand \bibfield  [0]{\@secondoftwo}%
\providecommand \translation [1]{[#1]}%
\providecommand \BibitemOpen [0]{}%
\providecommand \bibitemStop [0]{}%
\providecommand \bibitemNoStop [0]{.\EOS\space}%
\providecommand \EOS [0]{\spacefactor3000\relax}%
\providecommand \BibitemShut  [1]{\csname bibitem#1\endcsname}%
\let\auto@bib@innerbib\@empty
\bibitem [{\citenamefont {Lawrence}\ \emph {et~al.}(2016)\citenamefont {Lawrence}, \citenamefont {Ben-Amigi}, \citenamefont {Brown}, \citenamefont {Brown}, \citenamefont {Case}, \citenamefont {Chapman}, \citenamefont {Churilov}, \citenamefont {Colless}, \citenamefont {Content}, \citenamefont {Depoy}, \citenamefont {Evans}, \citenamefont {Farrell}, \citenamefont {Goodwin}, \citenamefont {Jacoby}, \citenamefont {Klauser}, \citenamefont {Kuehn}, \citenamefont {Lorente}, \citenamefont {Mali}, \citenamefont {Marshall}, \citenamefont {Muller}, \citenamefont {Nichani}, \citenamefont {Pai}, \citenamefont {Prochaska}, \citenamefont {Saunders}, \citenamefont {Schmidt}, \citenamefont {Shortridge}, \citenamefont {Staszak}, \citenamefont {Szentgyorgyi}, \citenamefont {Tims}, \citenamefont {Vuong}, \citenamefont {Waller},\ and\ \citenamefont {Zhelem}}]{Lawrence_2016}%
  \BibitemOpen
  \bibfield  {author} {\bibinfo {author} {\bibfnamefont {J.~S.}\ \bibnamefont {Lawrence}}, \bibinfo {author} {\bibfnamefont {S.}~\bibnamefont {Ben-Amigi}}, \bibinfo {author} {\bibfnamefont {D.~M.}\ \bibnamefont {Brown}}, \bibinfo {author} {\bibfnamefont {R.~A.}\ \bibnamefont {Brown}}, \bibinfo {author} {\bibfnamefont {S.}~\bibnamefont {Case}}, \bibinfo {author} {\bibfnamefont {S.}~\bibnamefont {Chapman}}, \bibinfo {author} {\bibfnamefont {V.}~\bibnamefont {Churilov}}, \bibinfo {author} {\bibfnamefont {M.}~\bibnamefont {Colless}}, \bibinfo {author} {\bibfnamefont {R.}~\bibnamefont {Content}}, \bibinfo {author} {\bibfnamefont {D.}~\bibnamefont {Depoy}}, \bibinfo {author} {\bibfnamefont {I.}~\bibnamefont {Evans}}, \bibinfo {author} {\bibfnamefont {T.}~\bibnamefont {Farrell}}, \bibinfo {author} {\bibfnamefont {M.}~\bibnamefont {Goodwin}}, \bibinfo {author} {\bibfnamefont {G.}~\bibnamefont {Jacoby}}, \bibinfo {author} {\bibfnamefont {U.}~\bibnamefont {Klauser}}, \bibinfo {author} {\bibfnamefont {K.}~\bibnamefont
  {Kuehn}}, \bibinfo {author} {\bibfnamefont {N.}~\bibnamefont {Lorente}}, \bibinfo {author} {\bibfnamefont {S.}~\bibnamefont {Mali}}, \bibinfo {author} {\bibfnamefont {J.}~\bibnamefont {Marshall}}, \bibinfo {author} {\bibfnamefont {R.}~\bibnamefont {Muller}}, \bibinfo {author} {\bibfnamefont {V.}~\bibnamefont {Nichani}}, \bibinfo {author} {\bibfnamefont {N.}~\bibnamefont {Pai}}, \bibinfo {author} {\bibfnamefont {T.}~\bibnamefont {Prochaska}}, \bibinfo {author} {\bibfnamefont {W.}~\bibnamefont {Saunders}}, \bibinfo {author} {\bibfnamefont {L.}~\bibnamefont {Schmidt}}, \bibinfo {author} {\bibfnamefont {K.}~\bibnamefont {Shortridge}}, \bibinfo {author} {\bibfnamefont {N.~F.}\ \bibnamefont {Staszak}}, \bibinfo {author} {\bibfnamefont {A.}~\bibnamefont {Szentgyorgyi}}, \bibinfo {author} {\bibfnamefont {J.}~\bibnamefont {Tims}}, \bibinfo {author} {\bibfnamefont {M.~V.}\ \bibnamefont {Vuong}}, \bibinfo {author} {\bibfnamefont {L.~G.}\ \bibnamefont {Waller}}, \ and\ \bibinfo {author} {\bibfnamefont {R.}~\bibnamefont
  {Zhelem}},\ }in\ \href@noop {} {\emph {\bibinfo {booktitle} {Ground-based and Airborne Instrumentation for Astronomy VI}}},\ \bibinfo {series} {Society of Photo-Optical Instrumentation Engineers (SPIE) Conference Series}, Vol.\ \bibinfo {volume} {9908},\ \bibinfo {editor} {edited by\ \bibinfo {editor} {\bibfnamefont {C.~J.}\ \bibnamefont {Evans}}, \bibinfo {editor} {\bibfnamefont {L.}~\bibnamefont {Simard}}, \ and\ \bibinfo {editor} {\bibfnamefont {T.}~\bibnamefont {H.}}}\ (\bibinfo {year} {2016})\ p.\ \bibinfo {pages} {99089O}\BibitemShut {NoStop}%
\bibitem [{\citenamefont {Bundy}\ \emph {et~al.}(2020{\natexlab{a}})\citenamefont {Bundy}, \citenamefont {Westfall}, \citenamefont {MacDonald}, \citenamefont {Kupke}, \citenamefont {Poppett}, \citenamefont {Miller}, \citenamefont {Lawrence}, \citenamefont {Lacombea}, \citenamefont {Yan}, \citenamefont {Goodwin}, \citenamefont {Kassis}, \citenamefont {O'Meara}, \citenamefont {Masters}, \citenamefont {Burchett}, \citenamefont {Williams}, \citenamefont {Rich}, \citenamefont {Villar}, \citenamefont {Sandford}, \citenamefont {Ting}, \citenamefont {Hinz}, \citenamefont {Schafer}, \citenamefont {Mandelbaum}, \citenamefont {Huang}, \citenamefont {Prochaska},\ and\ \citenamefont {Guhathakurta}}]{foboS}%
  \BibitemOpen
  \bibfield  {author} {\bibinfo {author} {\bibfnamefont {K.}~\bibnamefont {Bundy}}, \bibinfo {author} {\bibfnamefont {K.~B.}\ \bibnamefont {Westfall}}, \bibinfo {author} {\bibfnamefont {N.}~\bibnamefont {MacDonald}}, \bibinfo {author} {\bibfnamefont {R.}~\bibnamefont {Kupke}}, \bibinfo {author} {\bibfnamefont {C.}~\bibnamefont {Poppett}}, \bibinfo {author} {\bibfnamefont {T.~N.}\ \bibnamefont {Miller}}, \bibinfo {author} {\bibfnamefont {J.}~\bibnamefont {Lawrence}}, \bibinfo {author} {\bibfnamefont {C.~S.}\ \bibnamefont {Lacombea}}, \bibinfo {author} {\bibfnamefont {R.}~\bibnamefont {Yan}}, \bibinfo {author} {\bibfnamefont {M.}~\bibnamefont {Goodwin}}, \bibinfo {author} {\bibfnamefont {M.}~\bibnamefont {Kassis}}, \bibinfo {author} {\bibfnamefont {J.~M.}\ \bibnamefont {O'Meara}}, \bibinfo {author} {\bibfnamefont {D.~C.}\ \bibnamefont {Masters}}, \bibinfo {author} {\bibfnamefont {J.~N.}\ \bibnamefont {Burchett}}, \bibinfo {author} {\bibfnamefont {B.~F.}\ \bibnamefont {Williams}}, \bibinfo {author}
  {\bibfnamefont {R.~M.}\ \bibnamefont {Rich}}, \bibinfo {author} {\bibfnamefont {V.~A.}\ \bibnamefont {Villar}}, \bibinfo {author} {\bibfnamefont {N.}~\bibnamefont {Sandford}}, \bibinfo {author} {\bibfnamefont {Y.-S.}\ \bibnamefont {Ting}}, \bibinfo {author} {\bibfnamefont {P.}~\bibnamefont {Hinz}}, \bibinfo {author} {\bibfnamefont {C.}~\bibnamefont {Schafer}}, \bibinfo {author} {\bibfnamefont {R.}~\bibnamefont {Mandelbaum}}, \bibinfo {author} {\bibfnamefont {M.}~\bibnamefont {Huang}}, \bibinfo {author} {\bibfnamefont {J.~X.}\ \bibnamefont {Prochaska}}, \ and\ \bibinfo {author} {\bibfnamefont {P.}~\bibnamefont {Guhathakurta}},\ }in\ \href@noop {} {\emph {\bibinfo {booktitle} {Ground-based and Airborne Instrumentation for Astronomy VIII}}},\ Vol.\ \bibinfo {volume} {11447},\ \bibinfo {editor} {edited by\ \bibinfo {editor} {\bibfnamefont {C.~J.}\ \bibnamefont {Evans}}, \bibinfo {editor} {\bibfnamefont {J.~J.}\ \bibnamefont {Bryant}}, \ and\ \bibinfo {editor} {\bibfnamefont {K.}~\bibnamefont {Motohara}}},\
  \bibinfo {organization} {International Society for Optics and Photonics}\ (\bibinfo  {publisher} {SPIE},\ \bibinfo {year} {2020})\ p.\ \bibinfo {pages} {114471D}\BibitemShut {NoStop}%
\bibitem [{\citenamefont {Morita}\ \emph {et~al.}(2024)\citenamefont {Morita}, \citenamefont {Ishimura}, \citenamefont {Inoue}, \citenamefont {Nishimura}, \citenamefont {Takahashi}, \citenamefont {Tsuritani}, \citenamefont {Zoysa}, \citenamefont {Ishizaki}, \citenamefont {Suzuki},\ and\ \citenamefont {Noda}}]{Morita}%
  \BibitemOpen
  \bibfield  {author} {\bibinfo {author} {\bibfnamefont {R.}~\bibnamefont {Morita}}, \bibinfo {author} {\bibfnamefont {S.}~\bibnamefont {Ishimura}}, \bibinfo {author} {\bibfnamefont {T.}~\bibnamefont {Inoue}}, \bibinfo {author} {\bibfnamefont {K.}~\bibnamefont {Nishimura}}, \bibinfo {author} {\bibfnamefont {H.}~\bibnamefont {Takahashi}}, \bibinfo {author} {\bibfnamefont {T.}~\bibnamefont {Tsuritani}}, \bibinfo {author} {\bibfnamefont {M.~D.}\ \bibnamefont {Zoysa}}, \bibinfo {author} {\bibfnamefont {K.}~\bibnamefont {Ishizaki}}, \bibinfo {author} {\bibfnamefont {M.}~\bibnamefont {Suzuki}}, \ and\ \bibinfo {author} {\bibfnamefont {S.}~\bibnamefont {Noda}},\ }\href@noop {} {\bibfield  {journal} {\bibinfo  {journal} {Optica}\ }\textbf {\bibinfo {volume} {11}},\ \bibinfo {pages} {971} (\bibinfo {year} {2024})}\BibitemShut {NoStop}%
\bibitem [{\citenamefont {Inoue}\ \emph {et~al.}(2022)\citenamefont {Inoue}, \citenamefont {Yoshida}, \citenamefont {Gelleta}, \citenamefont {Izumi}, \citenamefont {Yoshida}, \citenamefont {Ishizaki}, \citenamefont {{De Zoysa}},\ and\ \citenamefont {Noda}}]{Noda}%
  \BibitemOpen
  \bibfield  {author} {\bibinfo {author} {\bibfnamefont {T.}~\bibnamefont {Inoue}}, \bibinfo {author} {\bibfnamefont {M.}~\bibnamefont {Yoshida}}, \bibinfo {author} {\bibfnamefont {J.}~\bibnamefont {Gelleta}}, \bibinfo {author} {\bibfnamefont {K.}~\bibnamefont {Izumi}}, \bibinfo {author} {\bibfnamefont {K.}~\bibnamefont {Yoshida}}, \bibinfo {author} {\bibfnamefont {K.}~\bibnamefont {Ishizaki}}, \bibinfo {author} {\bibfnamefont {M.}~\bibnamefont {{De Zoysa}}}, \ and\ \bibinfo {author} {\bibfnamefont {S.}~\bibnamefont {Noda}},\ }\href@noop {} {\bibfield  {journal} {\bibinfo  {journal} {Opt. Mater. Express}\ }\textbf {\bibinfo {volume} {13}},\ \bibinfo {pages} {3262} (\bibinfo {year} {2022})}\BibitemShut {NoStop}%
\bibitem [{\citenamefont {Han}\ \emph {et~al.}(2021)\citenamefont {Han}, \citenamefont {Smith}, \citenamefont {Ng}, \citenamefont {Anand}, \citenamefont {Katkus},\ and\ \citenamefont {Juodkazis}}]{ASEC2021-11143}%
  \BibitemOpen
  \bibfield  {author} {\bibinfo {author} {\bibfnamefont {M.}~\bibnamefont {Han}}, \bibinfo {author} {\bibfnamefont {D.}~\bibnamefont {Smith}}, \bibinfo {author} {\bibfnamefont {S.~H.}\ \bibnamefont {Ng}}, \bibinfo {author} {\bibfnamefont {V.}~\bibnamefont {Anand}}, \bibinfo {author} {\bibfnamefont {T.}~\bibnamefont {Katkus}}, \ and\ \bibinfo {author} {\bibfnamefont {S.}~\bibnamefont {Juodkazis}},\ }\href@noop {} {\bibfield  {journal} {\bibinfo  {journal} {Engineering Proceedings}\ }\textbf {\bibinfo {volume} {11}},\ \bibinfo {pages} {44} (\bibinfo {year} {2021})}\BibitemShut {NoStop}%
\bibitem [{\citenamefont {Haynes}\ and\ \citenamefont {McGrath}(2006)}]{haynes2006wide}%
  \BibitemOpen
  \bibfield  {author} {\bibinfo {author} {\bibfnamefont {R.}~\bibnamefont {Haynes}}\ and\ \bibinfo {author} {\bibfnamefont {A.}~\bibnamefont {McGrath}},\ }\href@noop {} {\bibfield  {journal} {\bibinfo  {journal} {New Astronomy Reviews}\ }\textbf {\bibinfo {volume} {50}},\ \bibinfo {pages} {329} (\bibinfo {year} {2006})}\BibitemShut {NoStop}%
\bibitem [{\citenamefont {Bundy}\ \emph {et~al.}(2020{\natexlab{b}})\citenamefont {Bundy}, \citenamefont {Westfall}, \citenamefont {MacDonald}, \citenamefont {Kupke}, \citenamefont {Poppett}, \citenamefont {Miller}, \citenamefont {Lawrence}, \citenamefont {Lacombea}, \citenamefont {Yan}, \citenamefont {Goodwin}, \citenamefont {Kassis}, \citenamefont {O'Meara}, \citenamefont {Masters}, \citenamefont {Burchett}, \citenamefont {Williams}, \citenamefont {Rich}, \citenamefont {Villar}, \citenamefont {Sandford}, \citenamefont {Ting}, \citenamefont {Hinz}, \citenamefont {Schafer}, \citenamefont {Mandelbaum}, \citenamefont {Huang}, \citenamefont {Prochaska},\ and\ \citenamefont {Guhathakurta}}]{Bundy_2020}%
  \BibitemOpen
  \bibfield  {author} {\bibinfo {author} {\bibfnamefont {K.}~\bibnamefont {Bundy}}, \bibinfo {author} {\bibfnamefont {K.}~\bibnamefont {Westfall}}, \bibinfo {author} {\bibfnamefont {N.}~\bibnamefont {MacDonald}}, \bibinfo {author} {\bibfnamefont {R.}~\bibnamefont {Kupke}}, \bibinfo {author} {\bibfnamefont {C.}~\bibnamefont {Poppett}}, \bibinfo {author} {\bibfnamefont {T.~N.}\ \bibnamefont {Miller}}, \bibinfo {author} {\bibfnamefont {J.}~\bibnamefont {Lawrence}}, \bibinfo {author} {\bibfnamefont {C.~S.}\ \bibnamefont {Lacombea}}, \bibinfo {author} {\bibfnamefont {R.}~\bibnamefont {Yan}}, \bibinfo {author} {\bibfnamefont {M.}~\bibnamefont {Goodwin}}, \bibinfo {author} {\bibfnamefont {M.}~\bibnamefont {Kassis}}, \bibinfo {author} {\bibfnamefont {J.~M.}\ \bibnamefont {O'Meara}}, \bibinfo {author} {\bibfnamefont {D.~C.}\ \bibnamefont {Masters}}, \bibinfo {author} {\bibfnamefont {J.~N.}\ \bibnamefont {Burchett}}, \bibinfo {author} {\bibfnamefont {B.~F.}\ \bibnamefont {Williams}}, \bibinfo {author} {\bibfnamefont
  {R.~M.}\ \bibnamefont {Rich}}, \bibinfo {author} {\bibfnamefont {V.~A.}\ \bibnamefont {Villar}}, \bibinfo {author} {\bibfnamefont {N.}~\bibnamefont {Sandford}}, \bibinfo {author} {\bibfnamefont {Y.-S.}\ \bibnamefont {Ting}}, \bibinfo {author} {\bibfnamefont {P.}~\bibnamefont {Hinz}}, \bibinfo {author} {\bibfnamefont {C.}~\bibnamefont {Schafer}}, \bibinfo {author} {\bibfnamefont {R.}~\bibnamefont {Mandelbaum}}, \bibinfo {author} {\bibfnamefont {M.}~\bibnamefont {Huang}}, \bibinfo {author} {\bibfnamefont {J.~X.}\ \bibnamefont {Prochaska}}, \ and\ \bibinfo {author} {\bibfnamefont {P.}~\bibnamefont {Guhathakurta}},\ }\bibfield  {booktitle} {\emph {\bibinfo {booktitle} {Ground-based and Airborne Instrumentation for Astronomy VIII}},\ }\href@noop {} {\bibfield  {journal} {\bibinfo  {journal} {SPIE}\ }\textbf {\bibinfo {volume} {11447}},\ \bibinfo {pages} {114471D} (\bibinfo {year} {2020}{\natexlab{b}})}\BibitemShut {NoStop}%
\bibitem [{\citenamefont {McLean}\ \emph {et~al.}(2010)\citenamefont {McLean}, \citenamefont {Ramsay},\ and\ \citenamefont {Takami}}]{Saunders_2010}%
  \BibitemOpen
  \bibinfo {editor} {\bibfnamefont {I.~S.}\ \bibnamefont {McLean}}, \bibinfo {editor} {\bibfnamefont {S.~K.}\ \bibnamefont {Ramsay}}, \ and\ \bibinfo {editor} {\bibfnamefont {H.}~\bibnamefont {Takami}},\ eds.,\ \href@noop {} {\emph {\bibinfo {title} {Ground-based and Airborne Instrumentation for Astronomy III}}},\ Vol.\ \bibinfo {volume} {7735},\ \bibinfo {organization} {International Society for Optics and Photonics}\ (\bibinfo  {publisher} {SPIE},\ \bibinfo {year} {2010})\BibitemShut {NoStop}%
\bibitem [{\citenamefont {{Navarro}}\ and\ \citenamefont {{Burge}}(2016)}]{Staszak_2016}%
  \BibitemOpen
  \bibinfo {editor} {\bibfnamefont {R.}~\bibnamefont {{Navarro}}}\ and\ \bibinfo {editor} {\bibfnamefont {J.~H.}\ \bibnamefont {{Burge}}},\ eds.,\ \href {\doibase 10.1117/12.2233583} {\emph {\bibinfo {title} {Advances in Optical and Mechanical Technologies for Telescopes and Instrumentation II}}},\ \bibinfo {series} {Society of Photo-Optical Instrumentation Engineers (SPIE) Conference Series}, Vol.\ \bibinfo {volume} {9912}\ (\bibinfo {year} {2016})\BibitemShut {NoStop}%
\bibitem [{\citenamefont {Kuehn}\ \emph {et~al.}(2014)\citenamefont {Kuehn}, \citenamefont {Lawrence}, \citenamefont {Brown}, \citenamefont {Case}, \citenamefont {Colless}, \citenamefont {Gers}, \citenamefont {Ireland}, \citenamefont {Richards}, \citenamefont {Saunders}, \citenamefont {Staszak}, \citenamefont {Tims},\ and\ \citenamefont {Waller}}]{Kuehn_2014}%
  \BibitemOpen
  \bibfield  {author} {\bibinfo {author} {\bibfnamefont {K.}~\bibnamefont {Kuehn}}, \bibinfo {author} {\bibfnamefont {J.}~\bibnamefont {Lawrence}}, \bibinfo {author} {\bibfnamefont {D.}~\bibnamefont {Brown}}, \bibinfo {author} {\bibfnamefont {S.}~\bibnamefont {Case}}, \bibinfo {author} {\bibfnamefont {M.}~\bibnamefont {Colless}}, \bibinfo {author} {\bibfnamefont {L.}~\bibnamefont {Gers}}, \bibinfo {author} {\bibfnamefont {M.}~\bibnamefont {Ireland}}, \bibinfo {author} {\bibfnamefont {S.}~\bibnamefont {Richards}}, \bibinfo {author} {\bibfnamefont {W.}~\bibnamefont {Saunders}}, \bibinfo {author} {\bibfnamefont {N.}~\bibnamefont {Staszak}}, \bibinfo {author} {\bibfnamefont {J.}~\bibnamefont {Tims}}, \ and\ \bibinfo {author} {\bibfnamefont {L.}~\bibnamefont {Waller}},\ }\href {\doibase doi:10.1117/12.2055677} {\bibfield  {journal} {\bibinfo  {journal} {Ground-Based and Airborne Instrumentation for Astronomy V}\ } (\bibinfo {year} {2014}),\ doi:10.1117/12.2055677}\BibitemShut {NoStop}%
\bibitem [{\citenamefont {Goodwin}\ \emph {et~al.}(2015)\citenamefont {Goodwin}, \citenamefont {Zheng}, \citenamefont {Lawrence}, \citenamefont {Richards},\ and\ \citenamefont {S.}}]{Goodwin_2015}%
  \BibitemOpen
  \bibfield  {author} {\bibinfo {author} {\bibfnamefont {M.}~\bibnamefont {Goodwin}}, \bibinfo {author} {\bibfnamefont {J.}~\bibnamefont {Zheng}}, \bibinfo {author} {\bibfnamefont {J.}~\bibnamefont {Lawrence}}, \bibinfo {author} {\bibfnamefont {S.}~\bibnamefont {Richards}}, \ and\ \bibinfo {author} {\bibfnamefont {L.-S.}\ \bibnamefont {S.}},\ }\href {\doibase https://doi.org/10.20353/K3T4CP1131608} {\enquote {\bibinfo {title} {Miniaturized {Shack-Hartmann} wavefront-sensors for elts},}\ } (\bibinfo {year} {2015})\BibitemShut {NoStop}%
\bibitem [{\citenamefont {Joglekar}\ \emph {et~al.}(2004)\citenamefont {Joglekar}, \citenamefont {Liu}, \citenamefont {Meyhöfer}, \citenamefont {Mourou},\ and\ \citenamefont {Hunt}}]{Joglekar2004}%
  \BibitemOpen
  \bibfield  {author} {\bibinfo {author} {\bibfnamefont {A.}~\bibnamefont {Joglekar}}, \bibinfo {author} {\bibfnamefont {H.}~\bibnamefont {Liu}}, \bibinfo {author} {\bibfnamefont {E.}~\bibnamefont {Meyhöfer}}, \bibinfo {author} {\bibfnamefont {G.}~\bibnamefont {Mourou}}, \ and\ \bibinfo {author} {\bibfnamefont {A.}~\bibnamefont {Hunt}},\ }\href@noop {} {\bibfield  {journal} {\bibinfo  {journal} {Proceedings of the National Academy of Sciences}\ }\textbf {\bibinfo {volume} {101}},\ \bibinfo {pages} {5856} (\bibinfo {year} {2004})}\BibitemShut {NoStop}%
\bibitem [{\citenamefont {Hua}\ \emph {et~al.}(2023)\citenamefont {Hua}, \citenamefont {Ren}, \citenamefont {Huang}, \citenamefont {Luan}, \citenamefont {Chen}, \citenamefont {Juodkazis},\ and\ \citenamefont {Sun}}]{23s2207968}%
  \BibitemOpen
  \bibfield  {author} {\bibinfo {author} {\bibfnamefont {J.-G.}\ \bibnamefont {Hua}}, \bibinfo {author} {\bibfnamefont {H.}~\bibnamefont {Ren}}, \bibinfo {author} {\bibfnamefont {J.}~\bibnamefont {Huang}}, \bibinfo {author} {\bibfnamefont {M.-L.}\ \bibnamefont {Luan}}, \bibinfo {author} {\bibfnamefont {Q.-D.}\ \bibnamefont {Chen}}, \bibinfo {author} {\bibfnamefont {S.}~\bibnamefont {Juodkazis}}, \ and\ \bibinfo {author} {\bibfnamefont {H.-B.}\ \bibnamefont {Sun}},\ }\href@noop {} {\bibfield  {journal} {\bibinfo  {journal} {Small}\ }\textbf {\bibinfo {volume} {19}},\ \bibinfo {pages} {2207968} (\bibinfo {year} {2023})}\BibitemShut {NoStop}%
\bibitem [{\citenamefont {Mamun}\ \emph {et~al.}(2021)\citenamefont {Mamun}, \citenamefont {Cadusch}, \citenamefont {Katkus}, \citenamefont {Juodkazis},\ and\ \citenamefont {Stoddart}}]{21olt107111}%
  \BibitemOpen
  \bibfield  {author} {\bibinfo {author} {\bibfnamefont {M.~A.~A.}\ \bibnamefont {Mamun}}, \bibinfo {author} {\bibfnamefont {P.~J.}\ \bibnamefont {Cadusch}}, \bibinfo {author} {\bibfnamefont {T.}~\bibnamefont {Katkus}}, \bibinfo {author} {\bibfnamefont {S.}~\bibnamefont {Juodkazis}}, \ and\ \bibinfo {author} {\bibfnamefont {P.~R.}\ \bibnamefont {Stoddart}},\ }\href@noop {} {\bibfield  {journal} {\bibinfo  {journal} {Optics \& Laser Technology}\ }\textbf {\bibinfo {volume} {141}},\ \bibinfo {pages} {107111} (\bibinfo {year} {2021})}\BibitemShut {NoStop}%
\bibitem [{\citenamefont {{Al Mamun}}\ \emph {et~al.}(2024)\citenamefont {{Al Mamun}}, \citenamefont {Katkus}, \citenamefont {Mahadevan-Jansen}, \citenamefont {Juodkazis},\ and\ \citenamefont {Stoddart}}]{24n1345}%
  \BibitemOpen
  \bibfield  {author} {\bibinfo {author} {\bibfnamefont {M.}~\bibnamefont {{Al Mamun}}}, \bibinfo {author} {\bibfnamefont {T.}~\bibnamefont {Katkus}}, \bibinfo {author} {\bibfnamefont {A.}~\bibnamefont {Mahadevan-Jansen}}, \bibinfo {author} {\bibfnamefont {S.}~\bibnamefont {Juodkazis}}, \ and\ \bibinfo {author} {\bibfnamefont {P.}~\bibnamefont {Stoddart}},\ }\href@noop {} {\bibfield  {journal} {\bibinfo  {journal} {Nanomaterials}\ }\textbf {\bibinfo {volume} {14}},\ \bibinfo {pages} {1345} (\bibinfo {year} {2024})}\BibitemShut {NoStop}%
\bibitem [{\citenamefont {Juodkazis}\ \emph {et~al.}(2008)\citenamefont {Juodkazis}, \citenamefont {Nishi},\ and\ \citenamefont {Misawa}}]{08pssrrl275}%
  \BibitemOpen
  \bibfield  {author} {\bibinfo {author} {\bibfnamefont {S.}~\bibnamefont {Juodkazis}}, \bibinfo {author} {\bibfnamefont {Y.}~\bibnamefont {Nishi}}, \ and\ \bibinfo {author} {\bibfnamefont {H.}~\bibnamefont {Misawa}},\ }\href@noop {} {\bibfield  {journal} {\bibinfo  {journal} {rrl - phys. stat. sol.}\ }\textbf {\bibinfo {volume} {2}},\ \bibinfo {pages} {275 } (\bibinfo {year} {2008})}\BibitemShut {NoStop}%
\bibitem [{\citenamefont {Juodkazis}\ \emph {et~al.}(2006)\citenamefont {Juodkazis}, \citenamefont {Nishimura}, \citenamefont {Misawa}, \citenamefont {Ebisui}, \citenamefont {Waki}, \citenamefont {Matsuo},\ and\ \citenamefont {Okada}}]{06am1361}%
  \BibitemOpen
  \bibfield  {author} {\bibinfo {author} {\bibfnamefont {S.}~\bibnamefont {Juodkazis}}, \bibinfo {author} {\bibfnamefont {K.}~\bibnamefont {Nishimura}}, \bibinfo {author} {\bibfnamefont {H.}~\bibnamefont {Misawa}}, \bibinfo {author} {\bibfnamefont {T.}~\bibnamefont {Ebisui}}, \bibinfo {author} {\bibfnamefont {R.}~\bibnamefont {Waki}}, \bibinfo {author} {\bibfnamefont {S.}~\bibnamefont {Matsuo}}, \ and\ \bibinfo {author} {\bibfnamefont {T.}~\bibnamefont {Okada}},\ }\href@noop {} {\bibfield  {journal} {\bibinfo  {journal} {Adv. Mat.}\ }\textbf {\bibinfo {volume} {18}},\ \bibinfo {pages} {1361 } (\bibinfo {year} {2006})}\BibitemShut {NoStop}%
\bibitem [{\citenamefont {Butkutė}\ \emph {et~al.}(2023)\citenamefont {Butkutė}, \citenamefont {Sirutkaitis}, \citenamefont {Gailevičius}, \citenamefont {Paipulas},\ and\ \citenamefont {Sirutkaitis}}]{etch}%
  \BibitemOpen
  \bibfield  {author} {\bibinfo {author} {\bibfnamefont {A.}~\bibnamefont {Butkutė}}, \bibinfo {author} {\bibfnamefont {R.}~\bibnamefont {Sirutkaitis}}, \bibinfo {author} {\bibfnamefont {D.}~\bibnamefont {Gailevičius}}, \bibinfo {author} {\bibfnamefont {D.}~\bibnamefont {Paipulas}}, \ and\ \bibinfo {author} {\bibfnamefont {V.}~\bibnamefont {Sirutkaitis}},\ }\href@noop {} {\bibfield  {journal} {\bibinfo  {journal} {Micromachines}\ }\textbf {\bibinfo {volume} {14}},\ \bibinfo {pages} {7} (\bibinfo {year} {2023})}\BibitemShut {NoStop}%
\bibitem [{\citenamefont {Bacigalupo}\ \emph {et~al.}(2018)\citenamefont {Bacigalupo}, \citenamefont {Goodwin}, \citenamefont {Brown}, \citenamefont {Brown}, \citenamefont {Case}, \citenamefont {Farrell}, \citenamefont {Kuehn},\ and\ \citenamefont {Lorente}}]{Bacigalupo_2018}%
  \BibitemOpen
  \bibfield  {author} {\bibinfo {author} {\bibfnamefont {C.}~\bibnamefont {Bacigalupo}}, \bibinfo {author} {\bibfnamefont {M.}~\bibnamefont {Goodwin}}, \bibinfo {author} {\bibfnamefont {R.}~\bibnamefont {Brown}}, \bibinfo {author} {\bibfnamefont {D.}~\bibnamefont {Brown}}, \bibinfo {author} {\bibfnamefont {S.}~\bibnamefont {Case}}, \bibinfo {author} {\bibfnamefont {T.}~\bibnamefont {Farrell}}, \bibinfo {author} {\bibfnamefont {K.}~\bibnamefont {Kuehn}}, \ and\ \bibinfo {author} {\bibfnamefont {N.}~\bibnamefont {Lorente}},\ }\bibfield  {booktitle} {\emph {\bibinfo {booktitle} {Advances in Optical and Mechanical Technologies for Telescopes and Instrumentation III}},\ }\href {\doibase 10.1117/12.2307057} {\bibfield  {journal} {\bibinfo  {journal} {SPIE}\ }\bibinfo {series} {Society of Photo-Optical Instrumentation Engineers (SPIE) Conference Series},\ \textbf {\bibinfo {volume} {10706}},\ \bibinfo {eid} {107065V} (\bibinfo {year} {2018})}\BibitemShut {NoStop}%
\bibitem [{\citenamefont {Zhu}\ and\ \citenamefont {Blackborow}(2011)}]{etendue}%
  \BibitemOpen
  \bibfield  {author} {\bibinfo {author} {\bibfnamefont {H.}~\bibnamefont {Zhu}}\ and\ \bibinfo {author} {\bibfnamefont {P.}~\bibnamefont {Blackborow}},\ }\href {https://www.energetiq.com/etendue-and-optical-throughput-calculations} {\enquote {\bibinfo {title} {Technical note \#002-2-14-2011: Etendue and optical throughput calculations},}\ } (\bibinfo {year} {2011}),\ \bibinfo {note} {energetiq Technology, Inc., Hamamatsu Photonics; last accessed 27 October 2024}\BibitemShut {NoStop}%
\bibitem [{\citenamefont {Lawrence}(2009)}]{inside}%
  \BibitemOpen
  \bibfield  {author} {\bibinfo {author} {\bibfnamefont {S.~A.}\ \bibnamefont {Lawrence}},\ }\href@noop {} {\enquote {\bibinfo {title} {Spherical lenses},}\ }\bibinfo {howpublished} {\url{http://www.physicsinsights.org/simple_optics_spherical_lenses-1.html}} (\bibinfo {year} {2009}),\ \bibinfo {note} {last accessed 5 Dec. 2024}\BibitemShut {NoStop}%
\bibitem [{\citenamefont {Wlodarczyk}(2011)}]{2011}%
  \BibitemOpen
  \bibfield  {author} {\bibinfo {author} {\bibfnamefont {K.}~\bibnamefont {Wlodarczyk}},\ }\emph {\bibinfo {title} {Surface deformation mechanisms in laser smoothing and micromachining of optical glasses}},\ \href@noop {} {Ph.D. thesis},\ \bibinfo  {school} {Heriot-Watt University, Edinburgh, UK} (\bibinfo {year} {2011})\BibitemShut {NoStop}%
\bibitem [{\citenamefont {Johnson}\ \emph {et~al.}(2014)\citenamefont {Johnson}, \citenamefont {Hultqvist},\ and\ \citenamefont {Bent}}]{C2_S2_P1}%
  \BibitemOpen
  \bibfield  {author} {\bibinfo {author} {\bibfnamefont {R.~W.}\ \bibnamefont {Johnson}}, \bibinfo {author} {\bibfnamefont {A.}~\bibnamefont {Hultqvist}}, \ and\ \bibinfo {author} {\bibfnamefont {S.~F.}\ \bibnamefont {Bent}},\ }\href@noop {} {\bibfield  {journal} {\bibinfo  {journal} {Materials Today}\ }\textbf {\bibinfo {volume} {17}} (\bibinfo {year} {2014})}\BibitemShut {NoStop}%
\bibitem [{\citenamefont {D.}\ \emph {et~al.}(2023)\citenamefont {D.}, \citenamefont {Galvanauskas}, \citenamefont {Gailevičius}, \citenamefont {Drazdys}, \citenamefont {Malinauskas},\ and\ \citenamefont {Grineviciute}}]{C2_S2_P4}%
  \BibitemOpen
  \bibfield  {author} {\bibinfo {author} {\bibfnamefont {A.}~\bibnamefont {D.}}, \bibinfo {author} {\bibfnamefont {K.}~\bibnamefont {Galvanauskas}}, \bibinfo {author} {\bibfnamefont {D.}~\bibnamefont {Gailevičius}}, \bibinfo {author} {\bibfnamefont {M.}~\bibnamefont {Drazdys}}, \bibinfo {author} {\bibfnamefont {M.}~\bibnamefont {Malinauskas}}, \ and\ \bibinfo {author} {\bibfnamefont {L.}~\bibnamefont {Grineviciute}},\ }\href@noop {} {\bibfield  {journal} {\bibinfo  {journal} {Nanomaterials}\ }\textbf {\bibinfo {volume} {13}} (\bibinfo {year} {2023})}\BibitemShut {NoStop}%
\bibitem [{\citenamefont {Pfeiffer}\ \emph {et~al.}(2017)\citenamefont {Pfeiffer}, \citenamefont {Schulz}, \citenamefont {Tünnermann},\ and\ \citenamefont {Szeghalmi}}]{C2_S2_P6}%
  \BibitemOpen
  \bibfield  {author} {\bibinfo {author} {\bibfnamefont {K.}~\bibnamefont {Pfeiffer}}, \bibinfo {author} {\bibfnamefont {U.}~\bibnamefont {Schulz}}, \bibinfo {author} {\bibfnamefont {A.}~\bibnamefont {Tünnermann}}, \ and\ \bibinfo {author} {\bibfnamefont {A.}~\bibnamefont {Szeghalmi}},\ }\href@noop {} {\bibfield  {journal} {\bibinfo  {journal} {Coatings}\ }\textbf {\bibinfo {volume} {7}} (\bibinfo {year} {2017})}\BibitemShut {NoStop}%
\bibitem [{\citenamefont {Myers}\ \emph {et~al.}(2021)\citenamefont {Myers}, \citenamefont {Throckmorton}, \citenamefont {Borrelli}, \citenamefont {O'Sullivan}, \citenamefont {Hatwar},\ and\ \citenamefont {George}}]{C2_S2_P20}%
  \BibitemOpen
  \bibfield  {author} {\bibinfo {author} {\bibfnamefont {T.~J.}\ \bibnamefont {Myers}}, \bibinfo {author} {\bibfnamefont {J.~A.}\ \bibnamefont {Throckmorton}}, \bibinfo {author} {\bibfnamefont {R.~A.}\ \bibnamefont {Borrelli}}, \bibinfo {author} {\bibfnamefont {M.}~\bibnamefont {O'Sullivan}}, \bibinfo {author} {\bibfnamefont {T.}~\bibnamefont {Hatwar}}, \ and\ \bibinfo {author} {\bibfnamefont {S.~M.}\ \bibnamefont {George}},\ }\href@noop {} {\bibfield  {journal} {\bibinfo  {journal} {Applied Surface Science}\ }\textbf {\bibinfo {volume} {569}},\ \bibinfo {pages} {150878} (\bibinfo {year} {2021})}\BibitemShut {NoStop}%
\bibitem [{\citenamefont {Bedi}\ and\ \citenamefont {Singh}(2016)}]{MRP}%
  \BibitemOpen
  \bibfield  {author} {\bibinfo {author} {\bibfnamefont {T.~S.}\ \bibnamefont {Bedi}}\ and\ \bibinfo {author} {\bibfnamefont {A.~K.}\ \bibnamefont {Singh}},\ }\href@noop {} {\bibfield  {journal} {\bibinfo  {journal} {Particulate Science and Technology}\ }\textbf {\bibinfo {volume} {34}},\ \bibinfo {pages} {412} (\bibinfo {year} {2016})}\BibitemShut {NoStop}%
\bibitem [{\citenamefont {Gamaly}\ and\ \citenamefont {Rode}(2018)}]{DieMet}%
  \BibitemOpen
  \bibfield  {author} {\bibinfo {author} {\bibfnamefont {E.~G.}\ \bibnamefont {Gamaly}}\ and\ \bibinfo {author} {\bibfnamefont {A.~V.}\ \bibnamefont {Rode}},\ }\href@noop {} {\bibfield  {journal} {\bibinfo  {journal} {Appl. Phys. A}\ }\textbf {\bibinfo {volume} {124}},\ \bibinfo {pages} {278} (\bibinfo {year} {2018})}\BibitemShut {NoStop}%
\bibitem [{\citenamefont {Hua}\ \emph {et~al.}(2020)\citenamefont {Hua}, \citenamefont {Ren}, \citenamefont {Jia}, \citenamefont {Tian}, \citenamefont {Wang}, \citenamefont {Juodkazis}, \citenamefont {Chen},\ and\ \citenamefont {Sun}}]{hua2020convex}%
  \BibitemOpen
  \bibfield  {author} {\bibinfo {author} {\bibfnamefont {J.-G.}\ \bibnamefont {Hua}}, \bibinfo {author} {\bibfnamefont {H.}~\bibnamefont {Ren}}, \bibinfo {author} {\bibfnamefont {A.}~\bibnamefont {Jia}}, \bibinfo {author} {\bibfnamefont {Z.-N.}\ \bibnamefont {Tian}}, \bibinfo {author} {\bibfnamefont {L.}~\bibnamefont {Wang}}, \bibinfo {author} {\bibfnamefont {S.}~\bibnamefont {Juodkazis}}, \bibinfo {author} {\bibfnamefont {Q.-D.}\ \bibnamefont {Chen}}, \ and\ \bibinfo {author} {\bibfnamefont {H.-B.}\ \bibnamefont {Sun}},\ }\href@noop {} {\bibfield  {journal} {\bibinfo  {journal} {Optics letters}\ }\textbf {\bibinfo {volume} {45}},\ \bibinfo {pages} {636} (\bibinfo {year} {2020})}\BibitemShut {NoStop}%
\bibitem [{\citenamefont {Zhao}\ \emph {et~al.}(2019{\natexlab{a}})\citenamefont {Zhao}, \citenamefont {Cheng}, \citenamefont {Chen}, \citenamefont {Yuan}, \citenamefont {Liao}, \citenamefont {Liu}, \citenamefont {Yang},\ and\ \citenamefont {Wang}}]{zhao2019formation}%
  \BibitemOpen
  \bibfield  {author} {\bibinfo {author} {\bibfnamefont {L.}~\bibnamefont {Zhao}}, \bibinfo {author} {\bibfnamefont {J.}~\bibnamefont {Cheng}}, \bibinfo {author} {\bibfnamefont {M.}~\bibnamefont {Chen}}, \bibinfo {author} {\bibfnamefont {X.}~\bibnamefont {Yuan}}, \bibinfo {author} {\bibfnamefont {W.}~\bibnamefont {Liao}}, \bibinfo {author} {\bibfnamefont {Q.}~\bibnamefont {Liu}}, \bibinfo {author} {\bibfnamefont {H.}~\bibnamefont {Yang}}, \ and\ \bibinfo {author} {\bibfnamefont {H.}~\bibnamefont {Wang}},\ }\href@noop {} {\bibfield  {journal} {\bibinfo  {journal} {International Journal of Extreme Manufacturing}\ }\textbf {\bibinfo {volume} {1}},\ \bibinfo {pages} {035001} (\bibinfo {year} {2019}{\natexlab{a}})}\BibitemShut {NoStop}%
\bibitem [{\citenamefont {Dudutis}\ \emph {et~al.}(2020)\citenamefont {Dudutis}, \citenamefont {Pipiras}, \citenamefont {Schwarz}, \citenamefont {Rung}, \citenamefont {Hellmann}, \citenamefont {Ra\v{c}iukaitis},\ and\ \citenamefont {Ge\v{c}ys}}]{S_LF_P25_Dudutis_2020}%
  \BibitemOpen
  \bibfield  {author} {\bibinfo {author} {\bibfnamefont {J.}~\bibnamefont {Dudutis}}, \bibinfo {author} {\bibfnamefont {J.}~\bibnamefont {Pipiras}}, \bibinfo {author} {\bibfnamefont {S.}~\bibnamefont {Schwarz}}, \bibinfo {author} {\bibfnamefont {S.}~\bibnamefont {Rung}}, \bibinfo {author} {\bibfnamefont {R.}~\bibnamefont {Hellmann}}, \bibinfo {author} {\bibfnamefont {G.}~\bibnamefont {Ra\v{c}iukaitis}}, \ and\ \bibinfo {author} {\bibfnamefont {P.}~\bibnamefont {Ge\v{c}ys}},\ }\href@noop {} {\bibfield  {journal} {\bibinfo  {journal} {Opt. Express}\ }\textbf {\bibinfo {volume} {28}},\ \bibinfo {pages} {5715} (\bibinfo {year} {2020})}\BibitemShut {NoStop}%
\bibitem [{\citenamefont {Zhao}\ \emph {et~al.}(2019{\natexlab{b}})\citenamefont {Zhao}, \citenamefont {Cheng}, \citenamefont {Chen}, \citenamefont {Yuan}, \citenamefont {Liao}, \citenamefont {Liu}, \citenamefont {Yang},\ and\ \citenamefont {Wang}}]{S_LF_P47_Zhao_2019}%
  \BibitemOpen
  \bibfield  {author} {\bibinfo {author} {\bibfnamefont {L.}~\bibnamefont {Zhao}}, \bibinfo {author} {\bibfnamefont {J.}~\bibnamefont {Cheng}}, \bibinfo {author} {\bibfnamefont {M.}~\bibnamefont {Chen}}, \bibinfo {author} {\bibfnamefont {X.}~\bibnamefont {Yuan}}, \bibinfo {author} {\bibfnamefont {W.}~\bibnamefont {Liao}}, \bibinfo {author} {\bibfnamefont {Q.}~\bibnamefont {Liu}}, \bibinfo {author} {\bibfnamefont {H.}~\bibnamefont {Yang}}, \ and\ \bibinfo {author} {\bibfnamefont {H.}~\bibnamefont {Wang}},\ }\href@noop {} {\bibfield  {journal} {\bibinfo  {journal} {International Journal of Extreme Manufacturing}\ }\textbf {\bibinfo {volume} {1}},\ \bibinfo {pages} {035001} (\bibinfo {year} {2019}{\natexlab{b}})}\BibitemShut {NoStop}%
\bibitem [{\citenamefont {Moon}\ \emph {et~al.}(2004)\citenamefont {Moon}, \citenamefont {Lee}, \citenamefont {Oh},\ and\ \citenamefont {Hutchinson}}]{MOON20043151}%
  \BibitemOpen
  \bibfield  {author} {\bibinfo {author} {\bibfnamefont {M.-W.}\ \bibnamefont {Moon}}, \bibinfo {author} {\bibfnamefont {K.-R.}\ \bibnamefont {Lee}}, \bibinfo {author} {\bibfnamefont {K.}~\bibnamefont {Oh}}, \ and\ \bibinfo {author} {\bibfnamefont {J.}~\bibnamefont {Hutchinson}},\ }\href@noop {} {\bibfield  {journal} {\bibinfo  {journal} {Acta Materialia}\ }\textbf {\bibinfo {volume} {52}},\ \bibinfo {pages} {3151} (\bibinfo {year} {2004})}\BibitemShut {NoStop}%
\bibitem [{\citenamefont {Smith}\ \emph {et~al.}(2023)\citenamefont {Smith}, \citenamefont {Ng}, \citenamefont {Tang}, \citenamefont {Katkus}, \citenamefont {Moraru},\ and\ \citenamefont {Juodkazis}}]{23p594}%
  \BibitemOpen
  \bibfield  {author} {\bibinfo {author} {\bibfnamefont {D.}~\bibnamefont {Smith}}, \bibinfo {author} {\bibfnamefont {S.}~\bibnamefont {Ng}}, \bibinfo {author} {\bibfnamefont {A.}~\bibnamefont {Tang}}, \bibinfo {author} {\bibfnamefont {T.}~\bibnamefont {Katkus}}, \bibinfo {author} {\bibfnamefont {D.}~\bibnamefont {Moraru}}, \ and\ \bibinfo {author} {\bibfnamefont {S.}~\bibnamefont {Juodkazis}},\ }\href@noop {} {\bibfield  {journal} {\bibinfo  {journal} {Photonics}\ }\textbf {\bibinfo {volume} {10}},\ \bibinfo {pages} {594} (\bibinfo {year} {2023})}\BibitemShut {NoStop}%
\bibitem [{\citenamefont {Polyanskiy}(2024)}]{nfo}%
  \BibitemOpen
  \bibfield  {author} {\bibinfo {author} {\bibfnamefont {M.~N.}\ \bibnamefont {Polyanskiy}},\ }\href {\doibase 10.1038/s41597-023-02898-2} {\enquote {\bibinfo {title} {Refractiveindex.info database of optical constants},}\ } (\bibinfo {year} {2024})\BibitemShut {NoStop}%
\bibitem [{\citenamefont {Zisman}(1964)}]{Zisman}%
  \BibitemOpen
  \bibfield  {author} {\bibinfo {author} {\bibfnamefont {W.}~\bibnamefont {Zisman}},\ }in\ \href@noop {} {\emph {\bibinfo {booktitle} {Contact Angle, Wettability, and Adhesion}}},\ Vol.~\bibinfo {volume} {43},\ \bibinfo {editor} {edited by\ \bibinfo {editor} {\bibfnamefont {F.}~\bibnamefont {Fowkes}}}\ (\bibinfo  {publisher} {Division of Colloid and Surface Chemistry},\ \bibinfo {year} {1964})\ Chap.~\bibinfo {chapter} {1}, pp.\ \bibinfo {pages} {1--51}\BibitemShut {NoStop}%
\bibitem [{\citenamefont {Eberhart}(1967)}]{Eberhart}%
  \BibitemOpen
  \bibfield  {author} {\bibinfo {author} {\bibfnamefont {J.~G.}\ \bibnamefont {Eberhart}},\ }\href@noop {} {\bibfield  {journal} {\bibinfo  {journal} {J. Physical Chemistry}\ }\textbf {\bibinfo {volume} {71}},\ \bibinfo {pages} {4125} (\bibinfo {year} {1967})}\BibitemShut {NoStop}%
\bibitem [{\citenamefont {Boyd}\ \emph {et~al.}(2012)\citenamefont {Boyd}, \citenamefont {Ebendorff-Heidepriem}, \citenamefont {Monro},\ and\ \citenamefont {Munch}}]{Boyd}%
  \BibitemOpen
  \bibfield  {author} {\bibinfo {author} {\bibfnamefont {K.}~\bibnamefont {Boyd}}, \bibinfo {author} {\bibfnamefont {H.}~\bibnamefont {Ebendorff-Heidepriem}}, \bibinfo {author} {\bibfnamefont {T.~M.}\ \bibnamefont {Monro}}, \ and\ \bibinfo {author} {\bibfnamefont {J.}~\bibnamefont {Munch}},\ }\href@noop {} {\bibfield  {journal} {\bibinfo  {journal} {Opt. Mater. Express}\ }\textbf {\bibinfo {volume} {2}},\ \bibinfo {pages} {1101} (\bibinfo {year} {2012})}\BibitemShut {NoStop}%
\bibitem [{\citenamefont {Cai}\ \emph {et~al.}(2021)\citenamefont {Cai}, \citenamefont {Jin}, \citenamefont {Kou}, \citenamefont {Zou}, \citenamefont {Xiao},\ and\ \citenamefont {Meng}}]{LW}%
  \BibitemOpen
  \bibfield  {author} {\bibinfo {author} {\bibfnamefont {J.}~\bibnamefont {Cai}}, \bibinfo {author} {\bibfnamefont {T.}~\bibnamefont {Jin}}, \bibinfo {author} {\bibfnamefont {J.}~\bibnamefont {Kou}}, \bibinfo {author} {\bibfnamefont {S.}~\bibnamefont {Zou}}, \bibinfo {author} {\bibfnamefont {J.}~\bibnamefont {Xiao}}, \ and\ \bibinfo {author} {\bibfnamefont {Q.}~\bibnamefont {Meng}},\ }\href@noop {} {\bibfield  {journal} {\bibinfo  {journal} {Langmuir}\ }\textbf {\bibinfo {volume} {37}},\ \bibinfo {pages} {1623} (\bibinfo {year} {2021})}\BibitemShut {NoStop}%
\bibitem [{CAS(2002)}]{CAS}%
  \BibitemOpen
  \href@noop {} {\enquote {\bibinfo {title} {Keck telescope and facility instrument guide},}\ }\bibinfo {howpublished} {\url{https://www2.keck.hawaii.edu/observing/kecktelgde/ktelinstupdate.pdf}} (\bibinfo {year} {2002}),\ \bibinfo {note} {last accessed 5 Dec. 2024}\BibitemShut {NoStop}%
\bibitem [{\citenamefont {Ryu}\ \emph {et~al.}(2024)\citenamefont {Ryu}, \citenamefont {Varapnickas}, \citenamefont {Gailevicius}, \citenamefont {Paipulas}, \citenamefont {Vilardell}, \citenamefont {Khajehsaeidimahabadi}, \citenamefont {Juodkazis}, \citenamefont {Morikawa},\ and\ \citenamefont {Malinauskas}}]{24sp059}%
  \BibitemOpen
  \bibfield  {author} {\bibinfo {author} {\bibfnamefont {M.}~\bibnamefont {Ryu}}, \bibinfo {author} {\bibfnamefont {S.}~\bibnamefont {Varapnickas}}, \bibinfo {author} {\bibfnamefont {D.}~\bibnamefont {Gailevicius}}, \bibinfo {author} {\bibfnamefont {D.}~\bibnamefont {Paipulas}}, \bibinfo {author} {\bibfnamefont {E.~P.}\ \bibnamefont {Vilardell}}, \bibinfo {author} {\bibfnamefont {Z.}~\bibnamefont {Khajehsaeidimahabadi}}, \bibinfo {author} {\bibfnamefont {S.}~\bibnamefont {Juodkazis}}, \bibinfo {author} {\bibfnamefont {J.}~\bibnamefont {Morikawa}}, \ and\ \bibinfo {author} {\bibfnamefont {M.}~\bibnamefont {Malinauskas}},\ }\href@noop {} {\bibfield  {journal} {\bibinfo  {journal} {SciPost Phys. Core}\ }\textbf {\bibinfo {volume} {7}},\ \bibinfo {pages} {059} (\bibinfo {year} {2024})}\BibitemShut {NoStop}%
\bibitem [{\citenamefont {Zhang}\ \emph {et~al.}(2023)\citenamefont {Zhang}, \citenamefont {Qian},\ and\ \citenamefont {Hong}}]{ZHANG}%
  \BibitemOpen
  \bibfield  {author} {\bibinfo {author} {\bibfnamefont {H.}~\bibnamefont {Zhang}}, \bibinfo {author} {\bibfnamefont {J.}~\bibnamefont {Qian}}, \ and\ \bibinfo {author} {\bibfnamefont {L.}~\bibnamefont {Hong}},\ }\href@noop {} {\bibfield  {journal} {\bibinfo  {journal} {Journal of Crystal Growth}\ }\textbf {\bibinfo {volume} {622}},\ \bibinfo {pages} {127402} (\bibinfo {year} {2023})}\BibitemShut {NoStop}%
\bibitem [{\citenamefont {Minardi}\ \emph {et~al.}(2021)\citenamefont {Minardi}, \citenamefont {Harris},\ and\ \citenamefont {Labadie}}]{Minardi_2021}%
  \BibitemOpen
  \bibfield  {author} {\bibinfo {author} {\bibfnamefont {S.}~\bibnamefont {Minardi}}, \bibinfo {author} {\bibfnamefont {R.}~\bibnamefont {Harris}}, \ and\ \bibinfo {author} {\bibfnamefont {L.}~\bibnamefont {Labadie}},\ }\href@noop {} {\bibfield  {journal} {\bibinfo  {journal} {The Astronomy and Astrophysics Review}\ }\textbf {\bibinfo {volume} {29}},\ \bibinfo {pages} {6} (\bibinfo {year} {2021})}\BibitemShut {NoStop}%
\bibitem [{\citenamefont {Astrauskyte}\ \emph {et~al.}(2023)\citenamefont {Astrauskyte}, \citenamefont {Galvanauskas}, \citenamefont {Gailevicius}, \citenamefont {Drazdys}, \citenamefont {Malinauskas},\ and\ \citenamefont {Grineviciute}}]{Astrauskyte}%
  \BibitemOpen
  \bibfield  {author} {\bibinfo {author} {\bibfnamefont {D.}~\bibnamefont {Astrauskyte}}, \bibinfo {author} {\bibfnamefont {K.}~\bibnamefont {Galvanauskas}}, \bibinfo {author} {\bibfnamefont {D.}~\bibnamefont {Gailevicius}}, \bibinfo {author} {\bibfnamefont {M.}~\bibnamefont {Drazdys}}, \bibinfo {author} {\bibfnamefont {M.}~\bibnamefont {Malinauskas}}, \ and\ \bibinfo {author} {\bibfnamefont {L.}~\bibnamefont {Grineviciute}},\ }\href@noop {} {\bibfield  {journal} {\bibinfo  {journal} {Nanomaterials}\ }\textbf {\bibinfo {volume} {13}},\ \bibinfo {pages} {2281} (\bibinfo {year} {2023})}\BibitemShut {NoStop}%
\bibitem [{\citenamefont {Grinevi\v{c}i\={u}t\.{e}}(2021)}]{Lina}%
  \BibitemOpen
  \bibfield  {author} {\bibinfo {author} {\bibfnamefont {L.}~\bibnamefont {Grinevi\v{c}i\={u}t\.{e}}},\ }\emph {\bibinfo {title} {Nanostructured Optical Coatings for the Manipulation of Laser Radiation}},\ \href@noop {} {Ph.D. thesis},\ \bibinfo  {school} {Vilnius University, Lithuania} (\bibinfo {year} {2021})\BibitemShut {NoStop}%
\bibitem [{\citenamefont {Mu}\ \emph {et~al.}(2023)\citenamefont {Mu}, \citenamefont {Smith}, \citenamefont {Katkus}, \citenamefont {Gailevičius}, \citenamefont {Malinauskas}, \citenamefont {Nishijima}, \citenamefont {Stoddart}, \citenamefont {Ruan}, \citenamefont {Ryu}, \citenamefont {Morikawa}, \citenamefont {Vasiliev}, \citenamefont {Lozovski}, \citenamefont {Moraru}, \citenamefont {Ng},\ and\ \citenamefont {Juodkazis}}]{24m798}%
  \BibitemOpen
  \bibfield  {author} {\bibinfo {author} {\bibfnamefont {H.}~\bibnamefont {Mu}}, \bibinfo {author} {\bibfnamefont {D.}~\bibnamefont {Smith}}, \bibinfo {author} {\bibfnamefont {T.}~\bibnamefont {Katkus}}, \bibinfo {author} {\bibfnamefont {D.}~\bibnamefont {Gailevičius}}, \bibinfo {author} {\bibfnamefont {M.}~\bibnamefont {Malinauskas}}, \bibinfo {author} {\bibfnamefont {Y.}~\bibnamefont {Nishijima}}, \bibinfo {author} {\bibfnamefont {P.~R.}\ \bibnamefont {Stoddart}}, \bibinfo {author} {\bibfnamefont {D.}~\bibnamefont {Ruan}}, \bibinfo {author} {\bibfnamefont {M.}~\bibnamefont {Ryu}}, \bibinfo {author} {\bibfnamefont {J.}~\bibnamefont {Morikawa}}, \bibinfo {author} {\bibfnamefont {T.}~\bibnamefont {Vasiliev}}, \bibinfo {author} {\bibfnamefont {V.}~\bibnamefont {Lozovski}}, \bibinfo {author} {\bibfnamefont {D.}~\bibnamefont {Moraru}}, \bibinfo {author} {\bibfnamefont {S.~H.}\ \bibnamefont {Ng}}, \ and\ \bibinfo {author} {\bibfnamefont {S.}~\bibnamefont {Juodkazis}},\ }\href@noop {} {\bibfield  {journal}
  {\bibinfo  {journal} {Micromachines}\ }\textbf {\bibinfo {volume} {14}},\ \bibinfo {pages} {798} (\bibinfo {year} {2023})}\BibitemShut {NoStop}%
\bibitem [{\citenamefont {Gu}(2000)}]{gu2000advanced}%
  \BibitemOpen
  \bibfield  {author} {\bibinfo {author} {\bibfnamefont {M.}~\bibnamefont {Gu}},\ }\href@noop {} {\emph {\bibinfo {title} {Advanced optical imaging theory}}},\ Vol.~\bibinfo {volume} {75}\ (\bibinfo  {publisher} {Springer Science \& Business Media},\ \bibinfo {year} {2000})\BibitemShut {NoStop}%
\bibitem [{\citenamefont {Wei}\ \emph {et~al.}(2021)\citenamefont {Wei}, \citenamefont {Cao}, \citenamefont {Lin}, \citenamefont {Mu}, \citenamefont {Liu}, \citenamefont {Yuan}, \citenamefont {Somekh},\ and\ \citenamefont {Jia}}]{wei2021high}%
  \BibitemOpen
  \bibfield  {author} {\bibinfo {author} {\bibfnamefont {S.}~\bibnamefont {Wei}}, \bibinfo {author} {\bibfnamefont {G.}~\bibnamefont {Cao}}, \bibinfo {author} {\bibfnamefont {H.}~\bibnamefont {Lin}}, \bibinfo {author} {\bibfnamefont {H.}~\bibnamefont {Mu}}, \bibinfo {author} {\bibfnamefont {W.}~\bibnamefont {Liu}}, \bibinfo {author} {\bibfnamefont {X.}~\bibnamefont {Yuan}}, \bibinfo {author} {\bibfnamefont {M.}~\bibnamefont {Somekh}}, \ and\ \bibinfo {author} {\bibfnamefont {B.}~\bibnamefont {Jia}},\ }\href@noop {} {\bibfield  {journal} {\bibinfo  {journal} {Photonics Research}\ }\textbf {\bibinfo {volume} {9}},\ \bibinfo {pages} {2454} (\bibinfo {year} {2021})}\BibitemShut {NoStop}%
\end{thebibliography}%
\end{document}